\documentclass[12pt,preprint]{aastex}
\usepackage{amsmath}

\let\url\relax
\usepackage[hypertex]{hyperref}
\makeatletter
\renewcommand{\fps@figure}{tp}
\makeatother
\setkeys{Gin}{width=\linewidth,height=0.9\textheight,keepaspectratio=true}

\newcommand{\elecd}{$n_{\rm e}$}
\newcommand{\elect}{$T_{\rm e}$}
\newcommand{\hb}{H$\beta$}

\newcommand{\foiii}{[\ion{O}{3}]}

\newcommand{\foii}{[\ion{O}{2}]}
\newcommand{\fsii}{[\ion{S}{2}]}
\newcommand{\fsiii}{[\ion{S}{3}]}
\newcommand{\fni}{[\ion{N}{1}]}
\newcommand{\fnii}{[\ion{N}{2}]}
\newcommand{\fariv}{[\ion{Ar}{4}]}
\newcommand{\fcliii}{[\ion{Cl}{3}]}

\newcommand{\ffeiii}{[\ion{Fe}{3}]}

\newcommand{\oii}{\ion{O}{2}}

\newcommand{\cii}{\ion{C}{2}}

\newcommand{\neii}{\ion{Ne}{2}}

\newcommand{\fariii}{[\ion{Ar}{3}]}

\newcommand{\hi}{\ion{H}{1}}
\newcommand{\hii}{\ion{H}{2}}

\newcommand{\hei}{\ion{He}{1}}
\newcommand{\heii}{\ion{He}{2}}

\newcommand{\ts}{\emph{$t^2$}}
\newcommand{\mc}{\multicolumn}

\shorttitle{On the abundance discrepancy problem in {\hii} regions}
\shortauthors{Garc\'{\i}a-Rojas \& Esteban}

\received{}
\begin{document}

\title{On the abundance discrepancy problem in {\hii} regions.\footnotemark{}}
\author{Jorge Garc\'{\i}a-Rojas} 
\affil{Instituto de Astrof{\'\i}sica de Canarias, E-38200, La Laguna, Tenerife, Spain}
\email{jogarcia@ll.iac.es}

\and

\author{C\'esar Esteban}
\affil{Instituto de Astrof{\'\i}sica de Canarias, E-38200, La Laguna, Tenerife, Spain}
\email{cel@ll.iac.es}

\begin{abstract} 
The origin of the abundance discrepancy --i.e. the fact that abundances derived from recombination lines 
are larger than those from collisionaly excited lines-- is one of the key problems in the physics of photoionized nebula. 
In this work, we analize and discuss data for a sample of
Galactic and extragalactic {\hii} regions where this abundance discrepancy has been determined. 
We find that the abundance discrepancy factor (ADF) is fairly constant  
and of the order of 2 in all the available sample of {\hii} regions. This is a rather different behaviour than that 
observed in planetary nebulae, where the ADF shows a much wider range of values.
We do not find correlations between the ADF and the O/H, O$^{++}$/H$^+$ ratios, the ionization degree, 
$T_e$(High), $T_e$(Low)/ $T_e$(High), FWHM,
and the effective temperature of the main ionizing stars within the observational uncertainties. 
These results indicate that whatever mechanism is producing  the abundance
discrepancy in {\hii} regions it does not substantially depend on those nebular parameters. On the contrary, 
the ADF seems to be slightly dependent on the
excitation energy, a fact that is consistent with the predictions of the classical temperature fluctuations paradigm. 
Finally, we obtain that $T_e$ values obtained from
{\oii} recombination lines in {\hii} regions are in agreement with those obtained from collisionally excited line 
ratios, a behaviour that is again different from that
observed in planetary nebulae. These similar temperature determinations are in contradiction with the predictions 
of the model based on the presence of chemically
inhomogeneous clumps but are consistent with the temperature fluctuations paradigm. 
We conclude that all the indications suggest that the physical mechanism responsible
of the abundance discrepancy in {\hii} regions and planetary nebulae are different.
\end{abstract}


\footnotetext[1]{Based on observations collected at the European Southern
Observatory, Chile, proposals number ESO 68.C-0149(A) and ESO 70.C-0008(A).}

\section{Introduction
\label{intro}}

{\hii} regions are essential tools for the study of the chemical composition and star formation in the Universe, 
specially in the 
extragalactic domain. It is necessary to be confident that our traditional techniques for deriving chemical 
abundances in 
ionized nebulae (based on the analysis of intensity ratios of colisionally excited lines, CELs) provide the real 
values. 

Forty years ago, \citet{peimbert67} characterized the temperature structure of a gaseous nebula to a second order 
by 
two parameters: the average temperature, $T_0$ and the mean square temperature fluctuation, {\ts}. At least, two 
independent 
methods of determining the electron temperature {\elect} of the nebula --with different weights 
in the high and low temperature regions-- are necessary to derive $T_0$ and {\ts}. 
In their observations, \citet{peimbert67} found that the temperature derived from the Balmer discontinuity, using 
the intensity ratio BJ/$I$({\hb}) where BJ is the value of the Balmer jump, was systematically lower than that 
derived from the 
{\foiii} 4363/5007 CELs ratio. This result has been corroborated by several subsequent works \citep[see e.g.][and 
references 
therein]{garciarojasetal06} and we will refer to this problem as the temperature discrepancy.

Later, \citet{peimbertcostero69} claimed that, in the presence of temperature variations over the observed 
volume of the nebula, the gaseous abundances derived from the analysis of intensity ratios of CELs 
are underestimated if these temperature fluctuations are not considered. 
This is due to the strong dependence of the intensity of those lines on the assumed {\elect}. 
This possibility is annoying, considering that the analysis of CELs is the standard method for deriving ionic 
abundances in 
ionized nebulae. On the other hand, intensity ratios of optical recombination lines (ORLs) are almost independent 
of the temperature structure 
of the nebula, and could be more appropiate to derive the ``real'' abundances of the nebulae. 
In the last years, several measurements of ORLs on {\hii} regions and planetary nebulae (PNe) have been done, 
and it has been found that 
abundance determinations from ORLs are systematically higher than those obtained using CELs, independently of the 
ion considered \citep[see e.g.][and references therein]{liu06,garciarojasetal06}. This is usually known as the 
abundance discrepancy problem.

Thanks to the fast technological progress --related to the development of large telescopes and better 
detectors-- and to the significant improvement in the understanding of the recombination theory for multielectronic 
atomic systems, many works have been dedicated to understand the origin of the abundance and temperature 
discrepancies. 
During the last years, it has been confirmed that both effects are real, and that they are not related neither to systematic 
biases in the 
observations nor in the basic atomic physics. In addition, it has been found that the two dichotomies can be 
intimately related 
\citep[see][]{liuetal00}. 

The high surface brightness of the brightest PNe has permitted to detect many ORLs of heavy elements in these 
objects. 
The pioneering work of \citet{wyse42} in NGC~7009, was the first in calling the attention on the different 
O$^{++}$ abundance obtained from {\oii} ORLs and {\foiii} CELs. Later, \citet{allermenzel45} and 
\citet{torrespeimbertpeimbert77} also remarked the excess of carbon observed in some PNe when calculating the 
C$^{++}$ abundance from {\cii} ORLs. Since the 
availability of the $IUE$ satellite, carbon abundances from UV CELs could be determined, obtaining abundances 
systematically smaller than those derived from ORLs \citep[see e.g.][]{harringtonetal80, kaler86, rolastasinska94, 
peimbertetal95}. During the last decade many efforts have been done to understand the physical conditions under 
which the different types of spectral lines are formed in PNe. In particular, the use of medium-large aperture 
telescopes, 
has allowed that the catalogue of detected {\oii} and {\cii} ORLs in PNe has increased spectacularly in the 
last years. The works dedicated to the study of the abundance discrepancy and/or the temperature structure in PNe 
are 
fairly numerous 
\citep[e.g.][and references therein]{garnettdinerstein01, mathisetal98, liuetal01, liuetal06, 
yliuetal04b, peimbertetal04, robertsontessigarnett05, tsamisetal04, wessonetal05}. 
The data collected in these works, indicate that the {\ts} parameter shows 
an extraordinary range of variation in PNe, reaching very high values in some objects. \citet{liuetal06}, found 
values 
of {\ts} $\sim$ 0.14 and 0.18 in Hf 2-2, from the abundance and temperature discrepancies, respectively. On the 
other hand, 
\citet{rubinetal02}, from $HST$ image and spectroscopy of PN NGC~7009, found that the {\elect} map of this nebula 
was practically uniform and that {\ts} throughout the nebula on the plane of the sky was small ({\ts}$_A$ = 
0.0035). In spite of the low {\ts} value found, it could not completely be discarded the existence of 
temperature fluctuations throughout the line of sight that could account for the observed abundance discrepancies in 
this object 
\citep[see][]{liuetal95}.

The abundance discrepancy is commonly quantified using the so-called abundance discrepancy factor (ADF), that 
is defined as:
\begin{equation}
\label{defadf}
{\rm ADF}({\rm X}^{+i})=({\rm X}^{+i}/{\rm H}^{+})_{\rm ORLs}/({\rm X}^{+i}/{\rm H}^{+})_{\rm CELs}. 
\end{equation}

\citet{liuetal01} found a correlation between the ADF and the difference between {\elect}({\foiii}) and 
{\elect}(Bac). Qualitatively, this correlation is consistent with the existence of temperature fluctuations, 
although these authors argue that abundance and temperature discrepancies are qualitatively better explained on the basis of 
a chemically inhomogenous or two-phase model.
\citet{liuetal00} analized the emission line spectrum of the PN NGC~6153, and concluded that the nebula contains a 
component of the 
ionized gas previously unknown --cold and very metal-rich--. Later works have supported 
this model, sustaining it mainly in the fact that the {\foiii} fine structure far infrared (FIR) lines at 
$\lambda\lambda$ 52, 
88 $\mu$m --which have very low excitation energies, of the order of 1000 K, and are almost insensible to the 
variations of {\elect}-- 
give ionic abundances comparable to the ones obtained from UV and optical CELs. These results are not compatible 
with the temperature fluctuations paradigm \citep[see][and references therein]{liu06}, but can be explained by a 
two-phase model (see below). 

The observational features that allow us to determine the {\ts} parameter (ORLs, {\hi} recombination continua) are 
much more 
difficult to be accurately measured in {\hii} regions than in PNe. 
The low surface brightness of most of these objects has caused 
that, until a few years ago, the {\hii} regions sample with {\ts} determinations available has been limited to a 
few Galactic {\hii} regions: the Orion nebula \citep{torrespeimbertetal80, osterbrocketal92, peimbertetal93, estebanetal98, 
rubinetal03}, 
M8 \citep{peimbertetal93b, estebanetal99b} and M17 \citep{peimbertetal92, estebanetal99a, tsamisetal03}, and to a few 
extragalactic {\hii} regions \citep{gonzalezdelgadoetal94, estebanetal02}. 
In the last four years, this sample has been extended with the works 
of \citet{tsamisetal03} on an additional Galactic {\hii} region (NGC~3576) and 3 extragalactic ones 
(1 in the Large Magellanic Cloud, LMC and 2 in the Small Magellanic Cloud, SMC); \citet{apeimbert03} on 30 Doradus; 
\citet{apeimbertetal05} on 2 NGC~6822 {\hii} regions; and \citet{lopezsanchezetal06} on the starburst galaxy 
NGC~5253. 
Aditionally, \citet{odelletal03} made a direct estimation of {\ts} in the plane of the sky of the Orion nebula from 
deep WFC/HST images, 
and making some geometrical considerations found a value of {\ts}=0.028$\pm$0.006, consistent with previous results \citep[{\ts} between 
0.020 and 0.030, see e.g.][and references therein]{estebanetal04}. 
Finally, our group has obtained and analyzed extremely deep 
echelle spectra of a sample of 8 Galactic {\hii} regions \citep{estebanetal04, garciarojasetal04, 
garciarojasetal05, garciarojasetal06, 
garciarojasetal06b}. In 4 of them (M16, M20, S~311 and NGC~3603), {\cii} and {\oii} ORLs were detected, for the 
first time; 
in other two regions (Orion nebula and NGC~3576) {\neii} ORLs were also detected for the first time. 

On the contrary, some puzzling results for the temperature fluctuations hypothesis have been recently found for some 
extragalactic 
{\hii} regions. On the one hand, \citet{hageleetal06} have measured {\elect}(Bac) in three compact {\hii} galaxies 
of the second data emission release of the SDSS\footnote{Sloan Digital Sky Survey} reobserved with 
the 4.2m William Herschel Telescope (WHT), finding {\ts}'s values close to zero for two of the objects, and a 
relatively large value ({\ts}$\sim$0.066$\pm$0.026) for the third.
On the other hand, \citet{gusevaetal06, gusevaetal07} have made a detailed study of a 
sample of 47 low metallicity {\hii} regions in blue compact dwarf (BCD) galaxies 
and and 69 emission-line galaxies from the data release 3 of the SDSS and M101. 
These authors have measured the Balmer and Paschen jumps with the purpose of 
determining {\elect}({\hi}). From Monte Carlo simulations, they fit the spectral energy distribution (SED) 
of the galaxies and find that the temperatures in the O$^{++}$ zone --derived from  
{\foiii} $\lambda\lambda$(4959 5007)/$\lambda$4363 ratio-- are, considering the dispersion of the data, equal to 
the temperature of 
the H$^+$ zone, obtained from fitting the Balmer jump and the SED. In any case, the 
dispersion of the data does not allow to discard small temperature fluctuations of about $\sim$ 3\% -- 5\%. 

\citet{tsamisetal03} proposed a chemically inhomogeneous 
scenario for {\hii} regions similar to that proposed for PNe, but only very recently a physically reasonable model 
has been drawn \citep{tsamispequignot05, stasinskaetal07}. 
This scenario implies the presence of temperature fluctuations in the nebulae, but its resulting abundance pattern is 
not identical to the one found under the standard temperature fluctuations paradigm \citep[see][]{stasinskaetal07}. 
Those authors postulate the presence 
of a gas component of high metallicity, and consequently low {\elect} embedded in a less dense ambient gas, 
with lower metallicity and therefore larger temperature. According to \citet{tsamispequignot05} and \citet{stasinskaetal07}, 
this denser component could come from  the resulting material of type II SNe that has not been mixed with the bulk of ISM and 
that is in pressure balance with the normal composition ambient gas \citep[an scenario originally proposed by][]{tenoriotagle96}.
According to this model, these inclusions would be responsible for most the ORLs emission and they would not emit in CELs due 
to their low temperature. 

In this paper, we present a global analysis of some results of our study of a sample of
{\hii} regions mainly focussed on discussing the properties and behavior of the abundance discrepancy problem in these objects.

The structure of this paper is as follows: in \S~\ref{datos} we briefly describe the data we have analized. 
In \S~\ref{adfvsparam} we explore the behavior of the abundance discrepancy factor (ADF) with respect to some 
nebular parameters. In \S~\ref{discusiont2} we present some observational arguments against 
the two phase model in {\hii} regions. Finally in \S~\ref{conclu} we briefly draw the conclusions of this work.

\section{The data
\label{datos}}

The observations were made in two runs: March 2002 and March 2003 with the Ultraviolet Visual Echelle Spectrograph 
(UVES) at the Very 
Large Telescope (VLT) Kueyen unit in Cerro Paranal Observatory (Chile). A detailed description of the instrument 
and telescope 
configuration, as well as of data reduction and analysis, can be found in previous papers 
\citep{estebanetal04, garciarojasetal04, garciarojasetal05, garciarojasetal06, garciarojasetal06b, 
lopezsanchezetal06}.

With the aim of presenting a homogeneous data set, we have recomputed the physical conditions and ionic and total 
abundances of some of our published results for some objects (NGC~3576, Orion nebula, 
NGC~3603) using the same atomic data and ionization correction factors (ICF) scheme than for the rest of the sample 
nebulae. 
Also, we have corrected some misprints ocurring in 
previous papers. In Table~\ref{plasma}, we present the physical conditions of the ionized gas 
computed for our sample; it can be noted that only 
small corrections in the adopted {\elect} and {\elecd} have been done with respect to the previously published 
ones. 
In Table~\ref{t2}, we present the estimated {\ts} values, which are essentially the same as the previously 
published ones. Also, in Table~\ref{rls} we present the compilation of the ionic abundances from ORLs.
In Tables~\ref{ionic} and~\ref{total}, we present the recomputed ionic and total abundances from CELs. We have re-calculated 
the electron 
density for the Orion nebula, NGC~3576 and NGC~3603. These new densities imply new values for the ionic abundances for the most density dependent 
ionic abundances i. e. O$^+$, S$^+$, Cl$^{++}$, Fe$^{3+}$ and Ar$^{+3}$. Changes in the 
adopted atomic data for O$^+$ and S$^{++}$ \citep[see][]{garciarojasetal05} have affected the determination of the 
ionic abundances and physical conditions of NGC~3576 and the Orion nebula. 
However, the combined effect of these corrections is small for NGC~3576 and NGC~3603 (less than 0.03 dex). 
For the Orion nebula the effect is somewhat more important, reaching a factor of about 0.1 dex for O$^+$.

Finally, we have re-computed He abundances for all the objects adopting a new set of ICFs: for the low ionization objects 
we have adopted the ICF by \citet{peimberttorrespeimbert77}, whereas for the high ionization ones we have adopted the ICF by 
\citet{peimbertetal92}.

\setcounter{table}{0}
\begin{deluxetable}{l@{\hspace{10pt}}l@{\hspace{10pt}}l@{\hspace{10pt}}l@{\hspace{10pt}}l@{\hspace{10pt}}l@{\hspace{10pt}}} 
\tabletypesize{\scriptsize}
\tablecaption{Plasma diagnostics.
\label{plasma}}
\tablewidth{0pt}
\tablehead{
\colhead{Diagnostic} & 
\colhead{Line}  & 
\colhead{M16}  & 
\colhead{M8}  & 
\colhead{M17}  & 
\colhead{M20}}  
\startdata 
{\elecd} (cm$^{-3}$)& [N\sc I] ($\lambda$5198)/($\lambda$5200)			& 1100$^{+750}_{-400}$ 	& 1600$^{+750}_{-470}$  & 1200$^{+1250}_{-500}$	     & 560$^{+340}_{-220}$     \\
& {\foii} ($\lambda$3726)/($\lambda$3729)					& 1050$\pm$250       	& 1800$\pm$800   	& 480$\pm$150	 	     & 240$\pm$70	       \\
& {\fsii} ($\lambda$6716)/($\lambda$6731)					& 1390$\pm$550	       	& 1600$\pm$450        	& 500$\pm$220	       	     & 320$\pm$130	       \\ 
& {\ffeiii} 									& 540$^{+>1000}_{-500}$	& 2600$\pm$1450       	& 430$^{+>1000}_{-400}$      & 560$\pm$390	       \\
& {\fcliii} ($\lambda$5518)/($\lambda$5538)					& 1370$\pm$1000	       	& 2100$\pm$700        	& 270$^{+630}_{-270}$        & 350$^{+780}_{-350}$     \\ 
& {\fariv} ($\lambda$4711)/($\lambda$4740) 					& \nodata	       	& 2450:$^{\rm d}$   	& $>$800		     & \nodata  	       \\
& {\elecd} (adopted) 								& 1120$\pm$220	       	& 1800$\pm$350		& 470$\pm$120	 	     & 270$\pm$60	       \\
& 										& 			& 			& 			     &  		       \\
$T_{\rm e}$ (K)& {\fnii} ($\lambda$6548+$\lambda$6583)/($\lambda$5755)$^{\rm a}$& 8450$\pm$270       	& 8470$\pm$180        	& 8950$\pm$380         	     & 8500$\pm$240	       \\
& {\fsii} ($\lambda$6716+$\lambda$6731)/($\lambda$4069+$\lambda$4076)		& 7520$\pm$430        	& 7220$\pm$300        	& 7100$\pm$750         	     & 6950 $\pm$350	       \\      
& {\foii} ($\lambda$3726+$\lambda$3729)/($\lambda$7320+$\lambda$7330)$^{\rm a}$ & 8260$\pm$400	       	& 8700$\pm$350        	& 8750$\pm$550         	     & 8275$\pm$400	       \\
& $T_{\rm e}$ (Low) 								& 8350$\pm$200	       	& 8500$\pm$150        	& 8870$\pm$300         	     & 8400$\pm$200	       \\
& {\foiii} ($\lambda$4959+$\lambda$5007)/($\lambda$4363) 			& 7650$\pm$250       	& 8090$\pm$140        	& 8020$\pm$170         	     & 7800$\pm$300	       \\
& {\fariii} ($\lambda$7136+$\lambda$7751)/($\lambda$5192)			& \nodata 	       	& 7550$\pm$420  	& 8380$\pm$570         	     & 8730$\pm$920$^{\rm b}$  \\      
& {\fsiii} ($\lambda$9069+$\lambda$9532)/($\lambda$6312) 			& 8430$\pm$450        	& 8600$\pm$300$^{\rm c}$& 8110$\pm$400         	     & 8300$\pm$400	       \\
& $T_{\rm e}$ (High) 								& 7850$\pm$220	       	& 8150$\pm$120        	& 8050$\pm$150         	     & 7980$\pm$250	       \\
& $T_{\rm e}$ (O~{\sc II+III}) 							& 8180$\pm$300	       	& 8570$\pm$200        	& 8200$\pm$200         	     & 8230$\pm$350	       \\
& {\heii} 									& 7300$\pm$350	       	& 7500$\pm$200        	& 7450$\pm$200         	     & 7650$\pm$300	       \\
& Balmer discontinuity								& 5450$\pm$820	       	& 7100$\pm$1100       	& \nodata 	       	     & 6000$\pm$1300	       \\ 
& Paschen discontinuity								& 7200$\pm$1300	       	& 7750$\pm$900        	& 6500$\pm$1000        	     & 5700$\pm$1300	       \\ \hline
\noalign{\smallskip}
& & \mc{1}{c}{NGC~3576} & \mc{1}{c}{ORION} & \mc{1}{c}{NGC~3603} & \mc{1}{c}{S~311} \\
\noalign{\smallskip}
\hline
\noalign{\smallskip}
{\elecd} (cm$^{-3}$)& {\fni} ($\lambda$5198)/($\lambda$5200)			& \nodata	       	& 1700$\pm$600  	 & 4000:$^{\rm d}$			& 590$^{+260}_{-200}$	\\
& {\foii} ($\lambda$3726)/($\lambda$3729)					& 1900$\pm$400         	& 6300$^{+2800}_{-1600}$ & 3000$\pm$1000	 	& 260$\pm$110         \\
& {\fsii} ($\lambda$6716)/($\lambda$6731)					& 1300$^{+500}_{-300}$ 	& 6500$^{+2000}_{-1200}$ & 4150$^{+3350}_{-1650}$	& 360$^{+140}_{-120}$   \\ 
& {\ffeiii} 									& 3100$\pm$ 1300       	& 9300$\pm$2700  	 & 1330:$^{\rm d}$			& 390$\pm$220		\\
& {\fcliii} ($\lambda$5518)/($\lambda$5538)					& 3500$^{+900}_{-700}$ 	& 9400$^{+1200}_{-700}$  & 5600$^{+3900}_{-2400}$	& 550$^{+650}_{-550}$   \\ 
& {\fariv} ($\lambda$4711)/($\lambda$4740) 					& 4500$^{+2600}_{-1800}$& 6800$^{+1100}_{-1000}$ & $\leq 4000$  		& \nodata		\\
& {\elecd} (adopted) 								& 2300$\pm$300         	& 7800$\pm$600		 & 3400$\pm$850 		& 310$\pm$80		\\
& 										&		       	&			 &				&			\\
$T_{\rm e}$ (K)&{\fnii} ($\lambda$6548+$\lambda$6583)/($\lambda$5755)$^{\rm a}$	& 8500$\pm$200         	& 10150$\pm$350 	 & 11050$\pm$800	 	& 9500$\pm$250        \\
& {\fsii} ($\lambda$6716+$\lambda$6731)/($\lambda$4069+$\lambda$4076)		& 8400$^{+350}_{-600}$ 	& 9050$\pm$800  	 & 11050$^{+3550}_{-2050}$	& 7200$^{+750}_{-600}$\\
& {\foii} ($\lambda$3726+$\lambda$3729)/($\lambda$7320+$\lambda$7330)$^{\rm a}$ & 8050$\pm$450         	& 8700$\pm$500	 	 & 12350$\pm$1250	 	& 9800$\pm$600        \\
& $T_{\rm e}$ (Low) 								& 8400$\pm$200         	& 9600$\pm$300		 & 11400$\pm$700	 	& 9550$\pm$250        \\
& {\foiii} ($\lambda$4959+$\lambda$5007)/($\lambda$4363) 			& 8500$\pm$50	       	& 8300$\pm$40		 & 9060$\pm$200 		& 9000$\pm$200        \\
& {\fariii} ($\lambda$7136+$\lambda$7751)/($\lambda$5192)			& 8600$^{+450}_{-350}$ 	& 8300$\pm$400  	 & \nodata			& 8800$^{+700}_{-850}$  \\
& {\fsiii} ($\lambda$9069+$\lambda$9532)/($\lambda$6312) 			& 8750$\pm$550         	& 9700$^{+800}_{-1200}$  & 8800 $\pm$500$^{\rm c}$	& 9300$\pm$350$^{\rm c}$\\
& $T_{\rm e}$ (High) 								& 8500$\pm$50	       	& 8320$\pm$40		 & 9030$\pm$200 		& 9050$\pm$200         \\
& $T_{\rm e}$ (O~{\sc II+III}) 							& 8500$\pm$50	       	& 8430$\pm$130		 & 9600$\pm$200 		& 9600$\pm$450         \\
& {\heii} 									& 6800$\pm$400         	& 7950$\pm$200		 & 8480$\pm$200 		& 8750$\pm$500         \\
& Balmer discontinuity								& 6650$\pm$750         	& 7900$\pm$600  	 & \nodata			& 9500$\pm$900         \\ 
& Paschen discontinuity								& 6700$\pm$900         	& 8100$\pm$1400 	 & 6900$\pm$1100	 	& 8700$\pm$1100	\\ 
\enddata
\tablenotetext{a}{The recombination contribution to the auroral lines has been taken into account.}
\tablenotetext{b}{The [Ar~{\sc iii}] $\lambda$7751 line is severely blended with a telluric line.}
\tablenotetext{c}{[S~{\sc iii}] $\lambda$9069 or $\lambda$9530 affected by atmospheric absorption bands.}
\tablenotetext{d}{Colons indicate very high uncertainties. These values has not been taken into account in the 
adopted average.}
\end{deluxetable}

\setcounter{table}{1}
\begin{deluxetable}{l@{\hspace{10pt}}c@{\hspace{10pt}}c@{\hspace{10pt}}c@{\hspace{10pt}}c@{\hspace{10pt}}} 
\tabletypesize{\scriptsize}
\tablecaption{{\ts} parameter.
\label{t2}}
\tablewidth{0pt}
\tablehead{
\colhead{Method} & 
\colhead{M16}  & 
\colhead{M8}  & 
\colhead{M17}  & 
\colhead{M20}}  
\startdata 
O$^{\rm +}$ (R/C)	& \nodata	      	& 0.031$\pm$0.017       & 0.109:  	        & 0.032$\pm$0.020       \\
O$^{\rm ++}$ (R/C)	& 0.046$\pm$0.007	& 0.045$\pm$0.005       & 0.034$\pm$0.005       & 0.038$\pm$0.016       \\
C$^{\rm ++}$ (R/C)	& \nodata		& 0.035$\pm$0.005 	& \nodata 	        & \nodata	       \\
Ne$^{\rm ++}$ (R/C)	& \nodata		& \nodata	 	& \nodata 	        & \nodata	       \\
He$^{\rm +}$ 		& 0.017$\pm$0.013	& 0.046$\pm$0.009 	& 0.027$\pm$0.014       & 0.017$\pm$0.010       \\
Bac/Pac--LEC 		& 0.045$\pm$0.014	& 0.022$\pm$0.015       & 0.035$\pm$0.021       & 0.049$\pm$0.019       \\ 
Adopted 		& {\bf 0.039$\pm$0.006}	& {\bf 0.040$\pm$0.004} & {\bf 0.033$\pm$0.005} & {\bf 0.029$\pm$0.007} \\ 
\hline
\noalign{\smallskip}
\mc{1}{c}{Method} & \mc{1}{c}{NGC~3576} & \mc{1}{c}{ORION} & \mc{1}{c}{NGC~3603} & \mc{1}{c}{S~311} \\
\noalign{\smallskip}
\hline
\noalign{\smallskip}
O$^{\rm +}$ (R/C)	& \nodata		   & 0.063$\pm$0.029       & \nodata		   & \nodata		   \\
O$^{\rm ++}$ (R/C)	& 0.038$\pm$0.006	   & 0.020$\pm$0.002	   & 0.042$\pm$0.009	   & 0.040$\pm$0.008     
\\
C$^{\rm ++}$ (R/C)	& \nodata		   & 0.039$\pm$0.011	   & \nodata		   & \nodata		   \\
Ne$^{\rm ++}$ (R/C)	& 0.036$^{+0.014}_{-0.024}$& 0.034$\pm$0.014	   & \nodata		   & \nodata		   \\
He$^{\rm +}$ 		& 0.023$\pm$0.019   	   & 0.022$\pm$0.002	   & 0.032$\pm$0.014	   & 0.034$\pm$0.010     
\\
Bac/Pac--LEC 		& 0.037$\pm$0.012      	   & 0.016$\pm$0.014	   & 0.056$\pm$0.023	   & 0.009:		   \\ 
Adopted 		& {\bf 0.038$\pm$0.009}    & {\bf 0.022$\pm$0.002} & {\bf 0.040$\pm$0.008} & {\bf 0.038$\pm$0.007}\\ 
\enddata
\end{deluxetable}

\setcounter{table}{2}
\begin{deluxetable}{l@{\hspace{10pt}}c@{\hspace{10pt}}c@{\hspace{10pt}}c@{\hspace{10pt}}c@{\hspace{10pt}}} 
\tabletypesize{\scriptsize}
\tablecaption{Heavy element ionic abundances from ORLs.
\label{rls}}
\tablewidth{0pt}
\tablehead{
 & \mc{4}{c}{12 + log(X$^{+i}$/H$^+$)}  \\ 
\noalign{\smallskip}
\hline
\noalign{\smallskip}
\colhead{Ion} & 
\colhead{M16}  & 
\colhead{M8}  & 
\colhead{M17}  & 
\colhead{M20}}  
\startdata 
O$^{\rm +}$ 		& \nodata	      	& 8.53$\pm$0.06         & \nodata  	        & 8.62$\pm$0.07       \\
O$^{\rm ++}$ 		& 8.30$\pm$0.04		& 8.23$\pm$0.02       	& 8.68$\pm$0.02         & 8.00$\pm$0.18       \\
C$^{\rm ++}$ 		& 8.40$\pm$0.03		& 8.30$\pm$0.02 	& 8.68$\pm$0.03 	& 8.18$\pm$0.05	       \\
Ne$^{\rm ++}$ 		& \nodata		& \nodata	 	& \nodata 	        & \nodata	       \\
\hline
\noalign{\smallskip}
\mc{1}{c}{Ion} & \mc{1}{c}{NGC~3576} & \mc{1}{c}{ORION} & \mc{1}{c}{NGC~3603} & \mc{1}{c}{S~311} \\
\noalign{\smallskip}
\hline
\noalign{\smallskip}
O$^{\rm +}$ 		& \nodata		   & \nodata       	   & \nodata		   & \nodata		   \\
O$^{\rm ++}$ 		& 8.62$\pm$0.05	   	   & 8.57$\pm$0.01	   & 8.71$\pm$0.05	   & 8.08$\pm$0.03     \\
C$^{\rm ++}$ 		& 8.45$\pm$0.06	   	   & 8.34$\pm$0.02	   & 8.48$\pm$0.07   	   & 8.00$\pm$0.04	   \\
Ne$^{\rm ++}$ 		& 7.90$\pm$0.14	 	   & 7.95$\pm$0.09	   & \nodata		   & \nodata		   \\
\enddata
\end{deluxetable}

\setcounter{table}{3}
\begin{deluxetable}{lcccccccc} 
\tabletypesize{\scriptsize}
\tablecaption{Heavy element ionic abundances from CELs.
\label{ionic}}
\tablewidth{0pt}
\tablehead{
 & \mc{8}{c}{12 + log(X$^{+i}$/H$^+$)}  \\ 
\noalign{\smallskip}
\hline
\noalign{\smallskip}
 & \mc{2}{c}{M16}  & 
\mc{2}{c}{M8}  & 
\mc{2}{c}{M17}  & 
\mc{2}{c}{M20} \\
\colhead{Ion} & 
\colhead{{\ts} = 0.000} & 
\colhead{{\ts} = 0.039} & 
\colhead{{\ts} = 0.000} & 
\colhead{{\ts} = 0.040} & 
\colhead{{\ts} = 0.000} & 
\colhead{{\ts} = 0.033} & 
\colhead{{\ts} = 0.000} & 
\colhead{{\ts} = 0.029}}  
\startdata 
N$^{0}$	   & 6.15$\pm$0.06 	& 6.33$\pm$0.07 &  5.81$\pm$0.05  	& 5.99$\pm$0.06 & 5.57$\pm$0.07 & 5.70$\pm$0.07		     	
& 5.90$\pm$0.07 & 6.03$\pm$0.08  \\
N$^{+}$	   & 7.71$\pm$0.05 	& 7.88$\pm$0.06 &  7.50$\pm$0.03  	& 7.67$\pm$0.04 & 6.82$\pm$0.10 & 6.94$\pm$0.10		     	
& 7.55$\pm$0.04 & 7.67$\pm$0.05  \\
O$^{0}$	   & 7.23$\pm$0.05 	& 7.40$\pm$0.06 &  6.40$\pm$0.03  	& 6.57$\pm$0.04 & 6.87$\pm$0.07 & 7.00$\pm$0.07		     	
& 6.60$\pm$0.05 & 6.72$\pm$0.06  \\
O$^{+}$	   & 8.47$\pm$0.08 	& 8.66$\pm$0.09 &  8.39$\pm$0.06  	& 8.58$\pm$0.07 & 7.84$\pm$0.09 & 7.98$\pm$0.09		     	
& 8.46$\pm$0.07 & 8.59$\pm$0.08  \\
O$^{++}$   &  7.85$\pm$0.07	& 8.18$\pm$0.10 &  7.86$\pm$0.03  	& 8.18$\pm$0.07 & 8.41$\pm$0.04 & 8.67$\pm$0.06		     	
& 7.67$\pm$0.08 & 7.90$\pm$0.10  \\
Ne$^{++}$  &  7.01$\pm$0.07	& 7.38$\pm$0.10 &  6.95$\pm$0.05  	& 7.30$\pm$0.07 & 7.64$\pm$0.04 & 7.93$\pm$0.07		     	
& 6.55$\pm$0.09 & 6.80$\pm$0.11  \\
S$^{+}$	   & 6.32$\pm$0.05 	& 6.49$\pm$0.06 &  5.93$\pm$0.04  	& 6.10$\pm$0.07 & 5.44$\pm$0.05 & 5.56$\pm$0.06		     	
& 6.17$\pm$0.05 & 6.29$\pm$0.06  \\
S$^{++}$   &  6.84$\pm$0.06	& 7.22$\pm$0.10 &  6.89$\pm$0.03  	& 7.25$\pm$0.07 & 6.90$\pm$0.04 & 7.19$\pm$0.06		     	
& 6.79$\pm$0.06 & 7.09$\pm$0.10  \\
Cl$^{+}$   &  4.77$\pm$0.05	& 4.91$\pm$0.07 &  4.53$\pm$0.04  	& 4.66$\pm$0.06 & 3.95$^{+0.09}_{-0.12}$& 
4.06$^{+0.09}_{-0.12}$& 4.75$\pm$0.05 & 4.85$\pm$0.07  \\
Cl$^{++}$  &  5.04$\pm$0.06	& 5.36$\pm$0.08 &  5.02$\pm$0.04  	& 5.32$\pm$0.06 & 5.04$\pm$0.04 & 5.29$\pm$0.06		     	
& 4.99$\pm$0.07 & 5.21$\pm$0.08  \\
Cl$^{3+}$  &  \nodata	    	& \nodata	   &  \nodata	     	& \nodata       & 3.15:	     	& 3.35:			     	
& \nodata       & \nodata        \\
Ar$^{++}$  &  6.25$\pm$0.05	& 6.53$\pm$0.08 &  6.21$\pm$0.03  	& 6.48$\pm$0.05 & 6.35$\pm$0.04 & 6.57$\pm$0.06		     	
& 6.17$\pm$0.06 & 6.36$\pm$0.08  \\
Ar$^{3+}$  &  3.89$\pm$0.22	& 4.23$\pm$0.23 &  3.69$\pm$0.09  	& 4.01$\pm$0.10 & 4.15$^{+0.12}_{-0.18}$& 
4.42$^{+0.13}_{-0.18}$& 4.01$\pm$0.18 & 4.24$\pm$0.19  \\
Fe$^{+}$   &  4.62:	   	& 4.78:	   	&  4.61: 	     	& 4.77:	     	& 4.05:	     	& 4.17:			     	& 
4.51:	     	& 4.62:	      	\\       
Fe$^{++}$  &  5.07$\pm$0.04	& 5.41$\pm$0.08 &  5.58$\pm$0.04  	& 5.91$\pm$0.06 & 5.24$\pm$0.06	& 5.51$\pm$0.08		     	
& 5.23$\pm$0.10 & 5.47$\pm$0.12  \\
Fe$^{3+}$  &  \nodata	   	& \nodata	&  \nodata	     	& \nodata       & \nodata       & \nodata  		     	
& \nodata       & \nodata        \\
\hline
\noalign{\smallskip} 
& \multicolumn{8}{c}{12 + log(X$^{+i}$/H$^+$)} \\
\noalign{\smallskip}
\hline
\noalign{\smallskip}
& \mc{2}{c}{NGC~3576} & \mc{2}{c}{ORION} & \mc{2}{c}{NGC~3603} & \mc{2}{c}{S~311} \\
\noalign{\smallskip}
Ion      & {\ts} = 0.000 	& {\ts} = 0.038 	& {\ts} = 0.000   & {\ts} = 0.022 		& {\ts} = 0.000 & {\ts} = 
0.040 & {\ts} = 0.000 & {\ts} = 0.038 				\\
\noalign{\smallskip} 
\hline
\noalign{\smallskip}
N$^{0}$	   &  \nodata	     &  \nodata 		& 5.69$\pm$0.06      & 5.76$\pm$0.06	     & 5.55$\pm$0.11 & 
5.65$\pm$0.11 & 5.74 $\pm$ 0.06       & 5.88 $\pm$ 0.07       \\
N$^{+}$	   & 7.09$\pm$0.06   & 7.25$\pm$0.07		& 6.97$\pm$0.09      & 7.03$\pm$0.09	     & 6.45$\pm$0.07 & 
6.55$\pm$0.07 & 7.25 $\pm$ 0.05       & 7.38 $\pm$ 0.06       \\
O$^{0}$	   & 6.35$\pm$0.04   & 6.51$\pm$0.06		& 6.22$\pm$0.05      & 6.28$\pm$0.05	     & 6.32$\pm$0.09 & 
6.42$\pm$0.09 & 6.74 $\pm$ 0.06       & 6.87 $\pm$ 0.06       \\
O$^{+}$	   & 8.21$\pm$0.07   & 8.38$\pm$0.08		& 7.93$\pm$0.15      & 8.00$\pm$0.15	     & 7.42$\pm$0.11 & 
7.52$\pm$0.11 & 8.26 $\pm$ 0.07       & 8.40 $\pm$ 0.08       \\
O$^{++}$   & 8.35$\pm$0.03   & 8.63$\pm$0.08		& 8.42$\pm$0.01      & 8.58$\pm$0.03	     & 8.42$\pm$0.05 & 
8.68$\pm$0.08 & 7.81 $\pm$ 0.04       & 8.05 $\pm$ 0.06       \\
Ne$^{++}$  & 7.61$\pm$0.09   & 7.91$\pm$0.12		& 7.67$\pm$0.07      & 7.84$\pm$0.07	     & 7.72$\pm$0.05 & 
8.00$\pm$0.08 & 6.81 $\pm$ 0.05       & 7.07 $\pm$ 0.07       \\
S$^{+}$	   & 5.75$\pm$0.08   & 5.91$\pm$0.09		& 5.44$\pm$0.06      & 5.50$\pm$0.06	     & 5.08$\pm$0.10 & 
5.17$\pm$0.10 & 6.03 $\pm$ 0.05       & 6.15 $\pm$ 0.06       \\
S$^{++}$   & 6.86$\pm$0.10   & 7.17$\pm$0.10		& 6.95$\pm$0.04      & 7.12$\pm$0.05	     & 6.83$\pm$0.04 & 
7.11$\pm$0.09 & 6.68 $\pm$ 0.07       & 6.95 $\pm$ 0.09       \\
Cl$^{+}$   & 4.13$\pm$0.08   & 4.26$\pm$0.08		& 3.80$\pm$0.11$^{\rm a}$& 3.85$\pm$0.11$^{\rm a}$& 
3.46$\pm$0.07 & 3.54$\pm$0.07 & 4.56 $\pm$ 0.05       & 4.67 $\pm$ 0.05       \\
Cl$^{++}$  & 4.95$\pm$0.06   & 5.21$^{+0.10}_{-0.08}$	& 5.13$\pm$0.02      & 5.28$\pm$0.02	     & 
5.07$\pm$0.05 & 5.31$\pm$0.08 & 4.85 $\pm$ 0.05       & 5.08 $\pm$ 0.05       \\
Cl$^{3+}$  & 3.21$\pm$0.07   & 3.42$\pm$0.09		& 3.66$\pm$0.12      & 3.78$\pm$0.12	     & 3.86$\pm$0.04 & 
4.06$\pm$0.07 & \nodata  	     & \nodata  	     \\ 
Ar$^{++}$  & 6.34$\pm$0.05   & 6.57$\pm$0.08		& 6.37$\pm$0.05      & 6.50$\pm$0.05	     & 6.35$\pm$0.04 & 
6.56$\pm$0.07 & 6.08 $\pm$ 0.04       & 6.28 $\pm$ 0.06       \\
Ar$^{3+}$  & 4.20$\pm$0.07   & 4.48$\pm$0.10		& 4.59$\pm$0.03      & 4.75$\pm$0.04	     & 4.85$\pm$0.06 & 
5.11$\pm$0.08 & 3.42$^{+0.18}_{-0.30}$& 3.66$^{+0.18}_{-0.30}$\\
Fe$^{+}$   & 4.61:	     & 4.76:  			& 4.54:	     	     & 4.60:  	     	     & 4.04:	     & 4.13:	     
& 4.25:  	     	     & 4.37:   		     \\ 
Fe$^{++}$  & 5.57$\pm$0.05   & 5.85$\pm$0.09		& 5.43$\pm$0.06      & 5.59$\pm$0.06	     & 5.24$\pm$0.06 & 
5.50$\pm$0.09 & 5.05 $\pm$ 0.06       & 5.30 $\pm$ 0.08       \\
Fe$^{3+}$  & 5.71$^{+0.17}_{-0.29}$& 5.95$^{+0.17}_{-0.29}$& 5.65$^{+0.19}_{-0.30}$ & 5.78$^{+0.19}_{-0.30}$& 
\nodata& \nodata	     & \nodata         	     & \nodata	 	     \\ 
\enddata
\tablenotetext{a}{These values were misprinted in \citet{estebanetal04}.}
\end{deluxetable}

\setcounter{table}{4}
\begin{deluxetable}{lcccccccc} 
\tabletypesize{\scriptsize}
\tablecaption{Total abundances.
\label{total}}
\tablewidth{0pt}
\tablehead{
 & \mc{8}{c}{12 + log(X$^{+i}$/H$^+$)}  \\ 
\noalign{\smallskip}
\hline
\noalign{\smallskip}
 & \mc{2}{c}{M16}  & 
\mc{2}{c}{M8}  & 
\mc{2}{c}{M17}  & 
\mc{2}{c}{M20} \\
& 
\colhead{{\ts} = 0.000} & 
\colhead{{\ts} = 0.039} & 
\colhead{{\ts} = 0.000} & 
\colhead{{\ts} = 0.040} & 
\colhead{{\ts} = 0.000} & 
\colhead{{\ts} = 0.033} & 
\colhead{{\ts} = 0.000} & 
\colhead{{\ts} = 0.029}}  
\startdata 
He	   	& 11.01$\pm$0.02& 10.97$\pm$0.02&  10.87$\pm$0.01	& 10.85$\pm$0.01      	& 10.97$\pm$0.01        & 
10.97$\pm$0.01       	& 10.95$\pm$0.06& 10.92$\pm$0.06\\
C$^{\rm a}$	& 8.85$\pm$0.10 & 8.85$\pm$0.10 &  8.61/8.69$\pm$0.09 	& 8.69/8.69$\pm$0.09    & 8.77$\pm$0.04         
& 8.77$\pm$0.04        	& 8.66$\pm$0.11 & 8.66$\pm$0.11 \\
N	   	& 7.84$\pm$0.06 & 8.07$\pm$0.12 &  7.72$\pm$0.03	& 7.96$\pm$0.06       	& 7.62$\pm$0.12	      	& 
7.87$\pm$0.13        	& 7.67$\pm$0.05 & 7.83$\pm$0.07 \\
O	   	& 8.56$\pm$0.07 & 8.78$\pm$0.07 &  8.51$\pm$0.05	& 8.73$\pm$0.05       	& 8.51$\pm$0.04	      	& 
8.75$\pm$0.05        	& 8.53$\pm$0.06 & 8.67$\pm$0.07 \\
O$^{\rm b}$	& 8.70$\pm$0.06 & 8.70$\pm$0.06 &  8.71$\pm$0.05	& 8.71$\pm$0.05       	& 8.89$\pm$0.04	      	& 
8.89$\pm$0.04        	& 8.71$\pm$0.09 & 8.71$\pm$0.09 \\
O$^{\rm c}$	& 8.81$\pm$0.07 & 8.81$\pm$0.07 &  8.71$\pm$0.04	& 8.71$\pm$0.04       	& 8.76$\pm$0.04	      	& 
8.76$\pm$0.04        	& 8.69$\pm$0.10 & 8.69$\pm$0.10 \\
Ne$^{\rm d}$   	& 7.86$\pm$0.15 & 8.08$\pm$0.17 &  7.81$\pm$0.12	& 8.03$\pm$0.13       	& 7.74$\pm$0.07	      	
& 8.01$\pm$0.09        	& 7.83$\pm$0.16 & 7.97$\pm$0.18 \\
S	   	& 6.96$\pm$0.05 & 7.29$\pm$0.08 &  6.94$\pm$0.03	& 7.28$\pm$0.06       	& 7.01$\pm$0.04	      	& 
7.33$\pm$0.06        	& 6.88$\pm$0.05 & 7.12$\pm$0.09 \\
Cl$^{\rm e}$	& 5.23$\pm$0.04 & 5.49$\pm$0.07 &  5.14$\pm$0.04	& 5.41$\pm$0.06       	& 5.08$\pm$0.04	    	
& 5.32$\pm$0.06	       	& 5.19$\pm$0.05 & 5.37$\pm$0.06 \\
Ar	   	& 6.70$\pm$0.06 & 6.84$\pm$0.08 &  6.52$\pm$0.05	& 6.69$\pm$0.06       	& 6.39$\pm$0.06	      	& 
6.59$\pm$0.07        	& 6.65$\pm$0.07 & 6.70$\pm$0.09 \\
Fe$^{\rm f}$	& 5.17$\pm$0.11 & 5.53$\pm$0.13 &  5.69$\pm$0.09	& 6.04$\pm$0.10       	& 5.87$\pm$0.13	      	
& 6.22$\pm$0.15        	& 5.31$\pm$0.13 & 5.56$\pm$0.15 \\
Fe$^{\rm g}$	& 5.20$\pm$0.06 & 5.51$\pm$0.07 &  5.62$\pm$0.04	& 5.94$\pm$0.06  	& 5.27$\pm$0.06		& 
5.52$\pm$0.08	      	& 5.31$\pm$0.09 & 5.52$\pm$0.10 \\
\hline
\noalign{\smallskip} 
& \multicolumn{8}{c}{12 + log(X$^{+i}$/H$^+$)} \\
\noalign{\smallskip}
\hline
\noalign{\smallskip}
& \mc{2}{c}{NGC~3576} & \mc{2}{c}{ORION} & \mc{2}{c}{NGC~3603} & \mc{2}{c}{S~311} \\
\noalign{\smallskip}
 & {\ts} = 0.000 	& {\ts} = 0.038 	& {\ts} = 0.000   & {\ts} = 0.022 		& {\ts} = 0.000 & {\ts} = 0.040 & 
{\ts} = 0.000 & {\ts} = 0.038 				\\
\noalign{\smallskip} 
\hline
\noalign{\smallskip}
He	   	& 10.97$\pm$0.03& 10.96$\pm$0.04& 10.95$\pm$0.004 & 10.95$\pm$0.004	& 10.99$\pm$0.01       	& 
10.99$\pm$0.01        & 10.99$\pm$0.02& 10.97$\pm$0.02\\
C$^{\rm a}$	& 8.61$\pm$0.08 & 8.61$\pm$0.08 & 8.43$\pm$0.04 & 8.43$\pm$0.04 	& 8.51$\pm$0.07        	& 
8.51$\pm$0.07         & 8.37$\pm$0.10 & 8.37$\pm$0.10\\
N	   	& 7.57$\pm$0.06 & 7.81$\pm$0.07 & 7.73$\pm$0.09 & 7.87$\pm$0.09 	& 7.62$\pm$0.13        	& 7.89$\pm$0.14         
& 7.43$\pm$0.06 & 7.61$\pm$0.07 \\
O	   	& 8.59$\pm$0.03 & 8.82$\pm$0.07 & 8.54$\pm$0.03 & 8.68$\pm$0.04 	& 8.46$\pm$0.05        	& 8.71$\pm$0.07         
& 8.39$\pm$0.05 & 8.56$\pm$0.06 \\
O$^{\rm b}$	& 8.74$\pm$0.06 & 8.74$\pm$0.06 & 8.71$\pm$0.03 & 8.71$\pm$0.03 	& \nodata	       	&  \nodata	       	
& \nodata	& \nodata       \\
O$^{\rm c}$	& 8.82$\pm$0.06 & 8.82$\pm$0.06 & 8.67$\pm$0.03 & 8.67$\pm$0.03 	& 8.72$\pm$0.05        	& 
8.72$\pm$0.05         & 8.57$\pm$0.05 & 8.57$\pm$0.05 \\
Ne$^{\rm d}$	& 7.85$\pm$0.10 & 8.11$\pm$0.12 & 7.79$\pm$0.07 & 7.94$\pm$0.07 	& 7.76$\pm$0.08        	& 
8.03$\pm$0.11         & 7.79$\pm$0.13 & 7.98$\pm$0.14 \\
S	   	& 6.92$\pm$0.10 & 7.23$\pm$0.10 & 7.04$\pm$0.04 & 7.23$\pm$0.04 	& 7.04$\pm$0.05        	& 7.36$\pm$0.08         
& 6.77$\pm$0.06 & 7.02$\pm$0.08 \\
Cl$^{\rm e}$	& 5.02$\pm$0.05 & 5.27$\pm$0.07 & 5.14$\pm$0.04	& 5.29$\pm$0.04 	& 5.11$\pm$0.05        	& 
5.34$\pm$0.07         & 5.03$\pm$0.06 & 5.22$\pm$0.07 \\
Ar	   	& 6.44$\pm$0.06 & 6.65$\pm$0.09 & 6.50$\pm$0.06 & 6.63$\pm$0.06 	& 6.37$\pm$0.06        	& 6.58$\pm$0.08         
& 6.43$\pm$0.06 & 6.56$\pm$0.07 \\
Fe$^{\rm f}$	& 5.89$\pm$0.21 & 6.23$\pm$0.10 & 5.96$\pm$0.16	& 6.18$\pm$0.14 	& 6.14$\pm$0.16        	& 
6.53$\pm$0.19         & 5.17$\pm$0.11 & 5.44$\pm$0.13 \\
Fe$^{\rm g}$	& 5.97$^{+0.11}_{-0.15}$ & 6.22$\pm$0.12& 5.88$\pm$0.12	& 6.02$\pm$0.12 & 5.27$\pm$0.06        	& 
5.52$\pm$0.09         & 5.11$\pm$0.06	& 5.35$\pm$0.08 \\
\enddata
\tablenotetext{a}{For M8 and Orion nebula, ICFs from C~{\sc ii}] $\lambda$2525 line/ICF from photoionization models 
by \citet{garnettetal99}.}
\tablenotetext{b}{O$^+$/H$^+$ and O$^{++}$/H$^+$ from O~{\sc i} and O~{\sc ii} ORLs.}
\tablenotetext{c}{O$^+$/H$^+$ from [O~{\sc ii}] CELs and {\ts}$>$0.00. O$^{++}$/H$^+$ from O~{\sc ii} ORLs.}
\tablenotetext{d}{ICF from \citet{peimbertcostero69} for M17, NGC~3576 and NGC~3603. ICF from 
\citet{torrespeimbertpeimbert77} for M16, M8, M20 and S~311.}
\tablenotetext{e}{Cl/H=Cl$^+$/H$^+$ + Cl$^{++}$/H$^+$ for M16, M8, M20, and S~311. Cl/H=Cl$^+$/H$^+$ + 
Cl$^{++}$/H$^{+}$ + Cl$^{+3}$/H$^+$ for M17, NGC~3576, Orion nebula and NGC~3603.}
\tablenotetext{f}{ICF from \citet{rodriguezrubin05} equation (2).}
\tablenotetext{g}{Fe/H=Fe$^+$/H$^+$ + Fe$^{++}$/H$^+$ for M16, M8, M17, M20, NGC~3603 and S~311. Fe/H=Fe$^+$/H$^+$ 
+ Fe$^{++}$/H$^{+}$ + Fe$^{+3}$/H$^+$ for NGC~3576 and Orion nebula.}
\end{deluxetable}

\section{Behavior of the ADF with respect to some nebular parameters.
\label{adfvsparam}} 

The different hypothesis proposed to explain the abundance discrepancy predict the existence or 
absence of correlations between the ADF and some nebular properties. 
Although these correlations have been exhaustively explored for PNe, 
there has not been a systematic study for {\hii} regions. 
The aim of this section is to analyze possible correlations between the ADF and different nebular properties 
in {\hii} regions (from our data and others available in the literature), with the purpose of discarding possible 
systematic 
errors and verifying the consistency of some hypothesis proposed to explain the abundance discrepancy. 

Hereafter, the ADF found by 
\citet{tsamisetal03} for the SMC N11B {\hii} region has been omitted in most of the figures and the discussion, 
because the ADF value for this region is much larger than for the rest of the {\hii} regions and may be 
considerably overestimated. For this object, \citet{tsamisetal03} 
made an attempt to correct the intensity of the RLs of multiplet 1 of {\oii} for the presence of stellar absorption features, mainly 
caused 
by B stars --that have a strong absorption {\oii} spectra-- in the area covered by the slit.  
In any case, this effect, which can be important in extragalactic objects, can only be corrected in the appropriate 
form if the spectral features of the stars are resolved, or if synthetic spectra are available, and this is not the 
case 
in this particular object. 
In fact, it is not possible to distinguish between the nebular emission and stellar absorption features in the 
spectra of \citet{tsamisetal03} of N11B. It is important to 
emphasize that this effect must be investigated whenever is possible to perform a suitable analysis when computing 
abundances from ORLs in extragalactic {\hii} regions. 

For all the fits, we have computed their Spearman's correlation coefficiens. These coefficients are the most appropriate ones
for non gaussian distributions, 
as well as the two-sided significance of its deviation from zero; the significance is a value in the interval [0.0, 1.0], and a small 
value indicates a significant correlation.

One important result of our data compilation for {\hii} regions is that the ADF is fairly constant for these objects and 
of the order of 2. This is a crucial difference with the behavior shown by this factor in PNe, where it can vary from values 
of 1 up to 20 (see \S~\ref{te_rls} for a further discussion and references).

\begin{figure}
\begin{center}
\epsscale{1.0}
\plotone{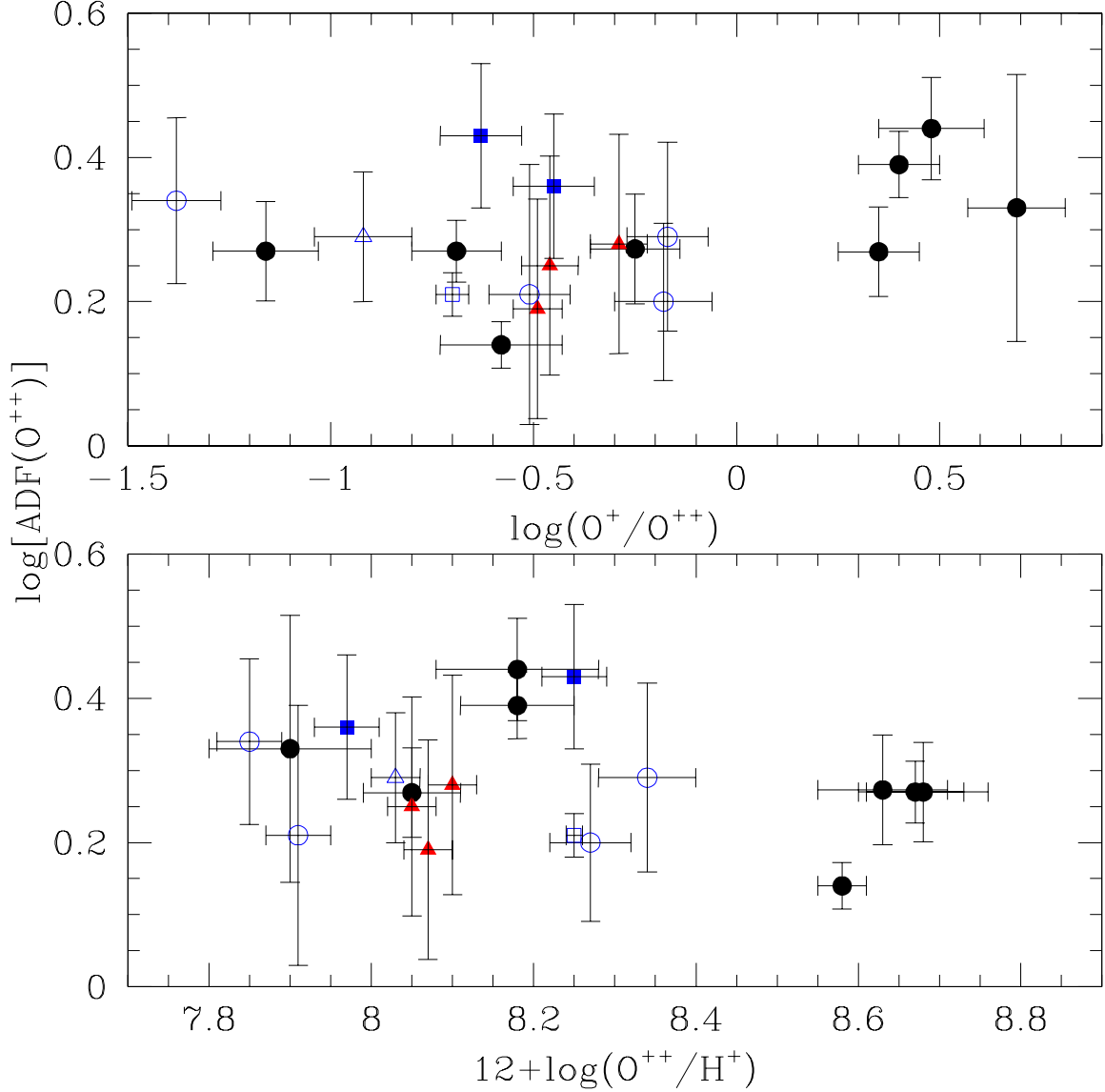} 
\figcaption{ADF(O$^{++}$) vs. the degree of ionization (upper panel) and O$^{++}$/H$^+$ determined from CELs (lower 
panel). 
The Galactic H~{\sc ii} regions of our sample 
are represented by filled dots; the rest of symbols represents data from the literature for 
extragalactic H~{\sc ii} regions: filled triangles: \citet{lopezsanchezetal06}; filled squares: 
\citet{tsamisetal03}; empty dots: \citet{estebanetal02}; empty triangles: \citet{apeimbertpeimbert05}; 
empty squares: \citet{apeimbert03}.
\label{adfoxy2}} 
\end{center} 
\end{figure} 

\begin{figure}
\begin{center}
\epsscale{1.0}
\plotone{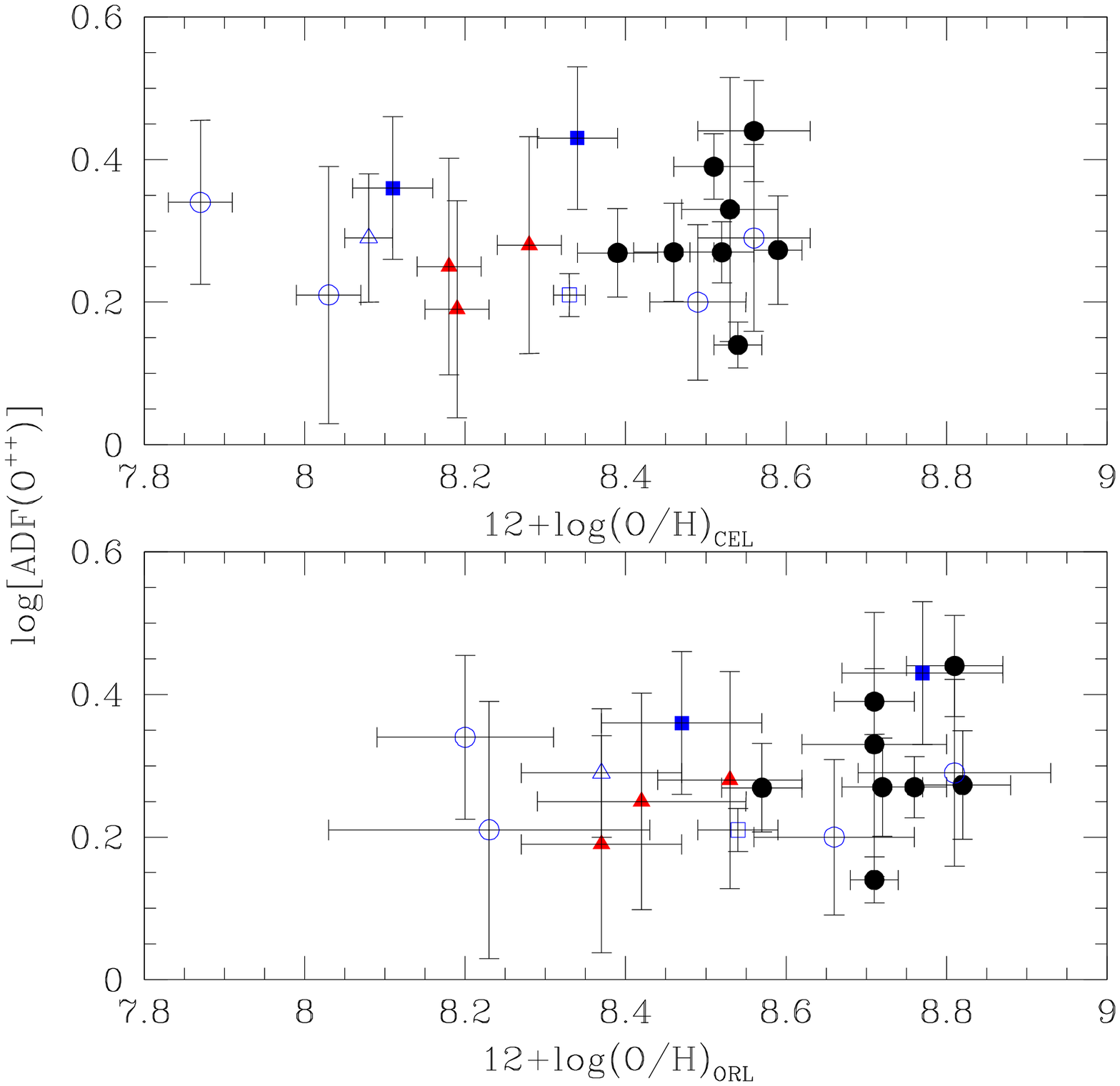} 
\figcaption{ADF(O$^{++}$) $vs.$ total O abundance derived from CELs (upper panel) 
and ORLs (lower panel). 
Symbols are the same that in Figure~\ref{adfoxy2}. 
\label{adf_Orl_cel}} 
\end{center} 
\end{figure}

\begin{figure}
\begin{center}
\epsscale{1.0}
\plotone{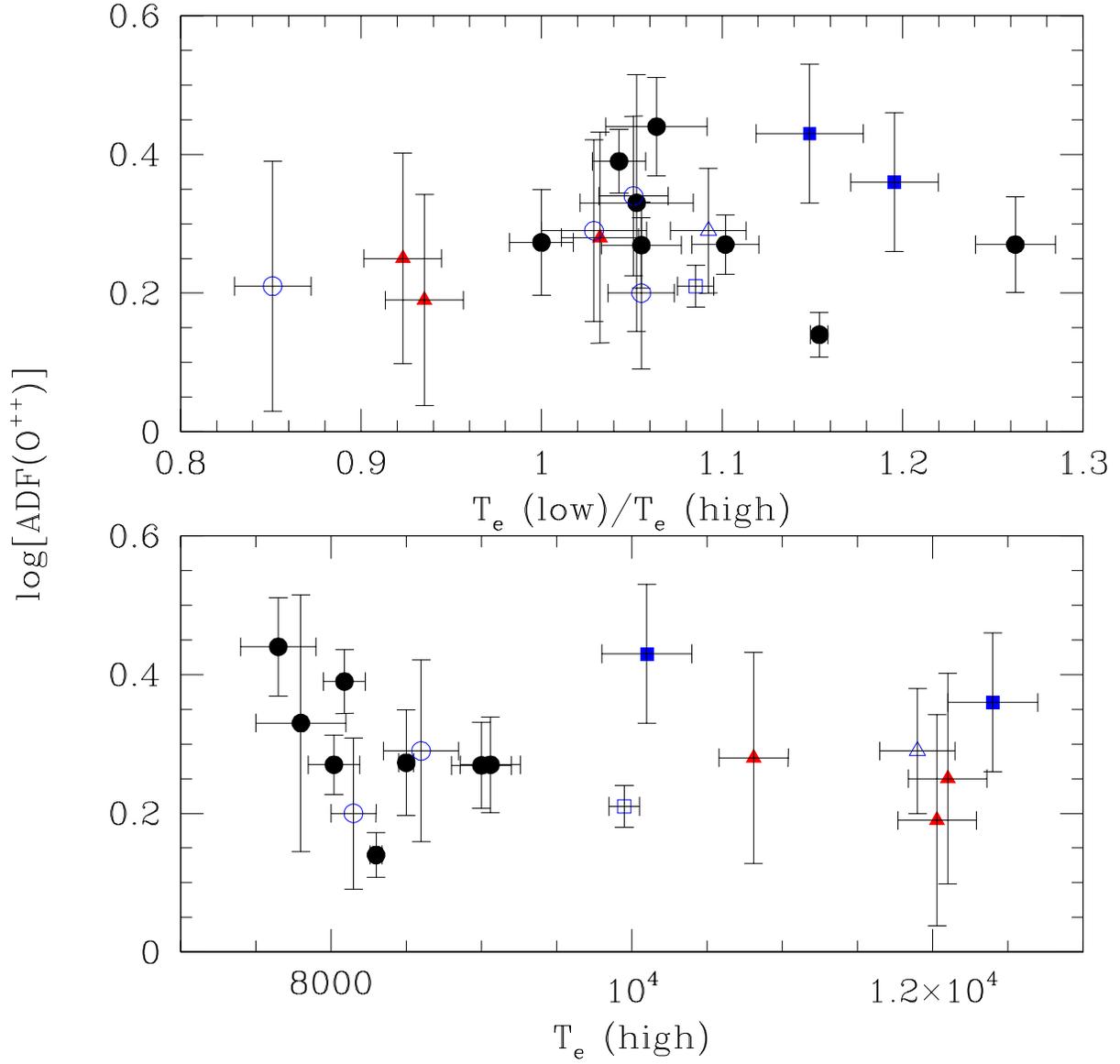} 
\figcaption{ADF(O$^{++}$) $vs.$ {\elect}(low)/{\elect}(high) (upper panel) and $vs.$ {\elect}(high) (lower panel). 
Symbols are the same that in Figure~\ref{adfoxy2}. 
\label{adftemp}} 
\end{center} 
\end{figure}

In Figure~\ref{adfoxy2} to ~\ref{adftemp} we 
have represented the ADF with respect to different nebular parameters, and in Table~\ref{coefcorr} we include the 
slope of the fits, their corresponding Spearman's coefficients, $r$, and the significance of the correlation, $P$, 
for each of the aforementioned diagrams. As it can be seen in 
Table~\ref{coefcorr} the correlation coefficients for all the fits are very low and they have quite poor statistical 
significance.
As in the rest of the cases, Figure~\ref{adfoxy2} shows that the ADF seems to 
be not related to O$^{++}$/H$^+$ nor O$^{+}$/O$^{++}$, at least within the observational uncertainties.
The metallicity dependency of the {\ts} parameter was suggested by \citet{garnett92}, who found 
that photoionization models could produce temperature fluctuations similar to the observed ones for 
nebulae with {\elect} $\leq$ 9000 K, increasing for colder (more metallic) nebulae. 
However, the observed {\ts} does not seem to depend on metallicity ,both for {\hii} regions and PNe. 
For PNe, several authors have found that the large range of abundance discrepancies found 
might be mainly due to 
the fact that ADF increases monotonically with metallicity \citep[see e.g.][]{liuetal00, yliuetal04b}. In the 
case of {\hii} regions, as it can be observed in Figure~\ref{adf_Orl_cel}, 
the ADF seems to remain relatively constant with the O abundance, determined either from CELs or ORLs, and 
within the uncertainties.
We have also represented the ADF $vs.$ the ratio between high and low ionization temperatures, finding that 
there is, again, no correlation within the observational errors (see Figure~\ref{adftemp} and Table~\ref{coefcorr}); 
this indicates that large scale 
variations of {\elect} 
due to natural temperature gradients throughout a nebula, do not seem to be related to the 
abundance discrepancy. As it can also been seen in Figure~\ref{adftemp}, we have verified that the ADF does not seem 
to depend on the assumed temperature, which discards systematic effects in the determination of abundances from CELs, 
at least at the precision of our data.

\setcounter{table}{5}
\begin{deluxetable}{cccc} 
\tablecaption{Parameters of the linear least-squares fits of the data represented 
in Figures~\ref{adfoxy2} to~\ref{adftemp}.
\label{coefcorr}}
\tablewidth{0pt}
\tablehead{
\colhead{ADF(O$^{++}$) vs. } & 
\colhead{Slope} & 
\colhead{$r$} &
\colhead{$P$}}
\startdata 
O$^+$/O$^{++}$		& 0.099		& 0.204  	& 0.411   \\
12+log(O$^{++}$)/H$^+$	& $-$0.164	& $-$0.198 	& 0.420   \\
12+log(O/H)$_{CEL}$	& $-$0.049	& 0.064  	& 0.768   \\
12+log(O/H)$_{ORL}$	& 0.088		& 0.283  	& 0.283   \\
{\elect}(Low)/{\elect}(High)& $-$0.379	& 0.225 	& 0.378   \\
{\elect}(High)		& 2.21$\times$10$^{-6}$& $-$0.113&0.646    \\
\enddata
\end{deluxetable}

\subsection{Testing the effect of gas kinematics.} 

\begin{figure}
\begin{center}
\epsscale{1.0}
\plotone{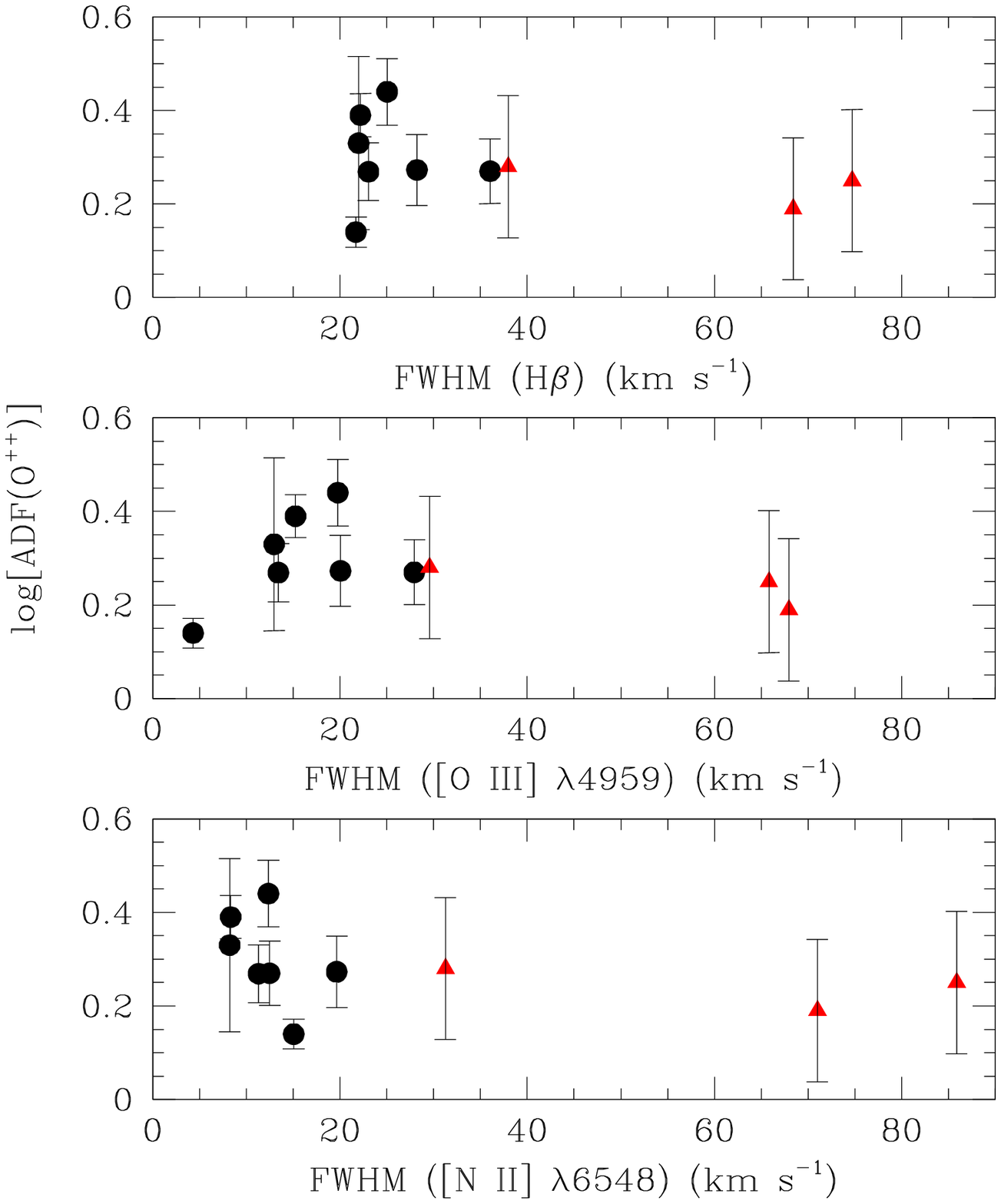} 
\figcaption{ADF(O$^{++}$) $vs.$ FWHM of {\hb} 
(upper panel), [O~{\sc iii}] $\lambda$4959 (medium panel) and [N~{\sc ii}] $\lambda$6548 (lower panel). 
Symbols are the same that in Figure~\ref{adfoxy2}. 
\label{adf_fwhm}} 
\end{center} 
\end{figure}

Several authors have proposed that mechanical energy deposited by shocks is a 
mechanism to increase the heating of nebulae and a possible source of temperature fluctuations 
\citep[e.g.][]{peimbert95, peimbertetal91}. 
One of the manifestations of the shocks consists of line broadening or line splitting due to the presence of 
different kinematical components. Hence, a way to investigate whether the 
ADF can be related to excitation by shocks is representing it with respect to the full width at half 
maximum (FWHM) of different emission lines. 
In Figure~\ref{adf_fwhm} we represent the ADF $vs.$ the FWHM of {\hb}, {\foiii} $\lambda$4959 and 
{\fnii} $\lambda$6548. In order to extend the FWHM baseline, we have introduced the points corresponding to 
3 slit positions of NGC~5253 analyzed by \citet{lopezsanchezetal06}. 
We have corrected the FWHM by the instrumental width. It is possible to 
note that FWHM are much larger for the extragalactic objects due to large scale gas movements, that are 
not observed in the Galactic {\hii} regions. With the available data, we cannot find a clear relation 
between the ADF and the represented FWHMs. It is obvious that the paucity of points with large FWHM does not 
permit to conclude any strong statement about this possible relation. Therefore, it is necessary to increase 
the number of extragalactic objects with good determinations of both quantities for making a suitable exploration.

\subsection{Testing the effect of dust grain heating.} 

Another possible effect that can affect the ADF was proposed by \citet{stasinskaszczerba01} and consists of 
photoelectric heating of dust grains in the zones near the ionizing star of a PN that increase {\elect} in the 
inner zones producing large {\ts} values. This is the case if the nebula is inhomogeneous in density, because the 
fluctuation in the ionization parameter induced by the variation of density, produce {\ts} due to differential dust and hydrogen 
heating. In the case of an homogeneous density, dust produces a temperature gradient but not real temperature fluctuations 
\citep{stasinskaszczerba01}. 
A way to verify if the effect of dust grains is significant in the observed ADF is to look for a correlation between the effective 
temperature, $T_{eff}$, of the ionizing star of the nebula --that gives an indication of the hardness of the radiation field-- 
and the ADF found in the associated nebula. 
\citet{robertsontessigarnett05} obtained that for PNe, the ADF was not correlated with $T_{eff}$ of the 
ionizing star. For {\hii} regions this is a complicated task because they are 
generally ionized by an OB association. In Table~\ref{adf_tipo} we include the $T_{eff}$ of the main ionizing star 
(the hottest one) of each studied {\hii} region, and it is clear that such quantity seems to be not correlated with the ADF of the 
associated nebula.

\setcounter{table}{6}
\begin{deluxetable}{cccc} 
\tablecaption{Comparison of the ADF with the spectral type of the main ionizing sources of each of our sample 
{\hii} regions.
\label{adf_tipo}}
\tablewidth{0pt}
\tablehead{
\colhead{Object$^{\rm a}$} & 
\colhead{Star/Cluster} & 
\colhead{Spectral type$^{\rm b}$} & 
\colhead{ADF(O$^{++}$)}}  
\startdata 
NGC~3603& NGC~3603/HST-38/40/A2/16& O3V		& 1.9  \\
	& NGC~3603/HST-42/A3	& O2-3III	&   	\\
M16     & HD168075 		& O4V		& 2.8  \\
M17	& Kleinmann's star	& O4V		& 2.1  \\
M20	& HD164492		& O6V		& 2.1  \\
Ori\'on & $\theta^1$Ori C	& O6peV		& 1.4  \\
S~311	& HD~64315		& O6eV		& 1.9	\\
M8   	& H36 			& O7.5V		& 2.0  \\
NGC~3576& Obscured cluster 	& OB association & 1.8  \\
\enddata
\tablenotetext{a}{Ordered from earlier to later spectral type.}
\tablenotetext{b}{From the Galactic O star catalog by \citet{maizapellanizetal04}.}
\end{deluxetable}

\subsection{The ADF and recombination continuum/CELs temperature difference.}

\begin{figure}
\begin{center}
\epsscale{1.0}
\plotone{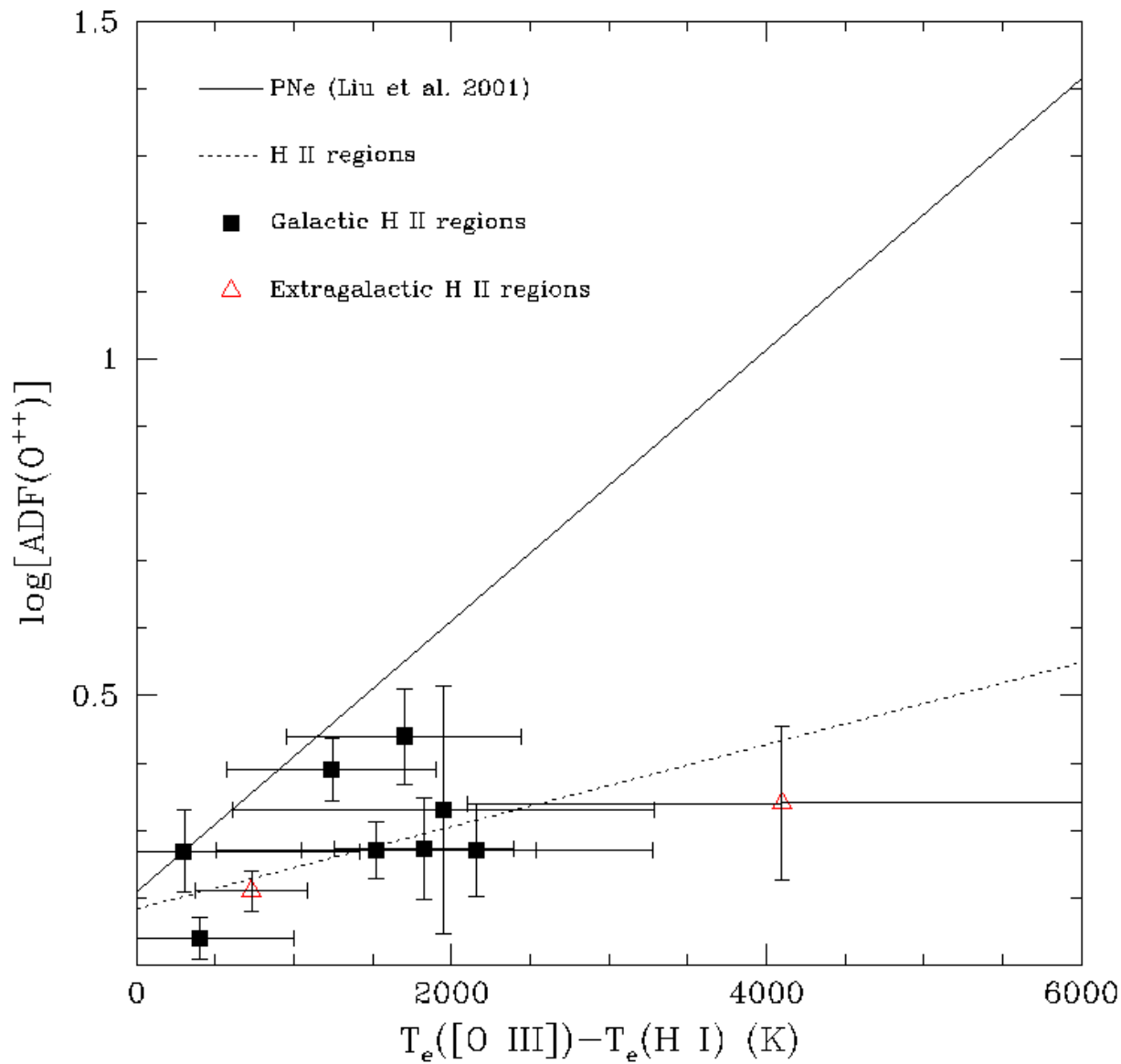} 
\figcaption{ADF(O$^{++}$) $vs.$ the difference between {\elect}(CELs) and {\elect}(H~{\sc i}). 
Solid line corresponds to the fit found by \citet{liuetal00} for PNe, whereas pointed line corresponds 
to the fit to our data of {\hii} regions. We have added the data of two extragalactic H~{\sc ii} regions: 30 
Doradus 
\citep{apeimbert03} and NGC~2363 \citep{gonzalezdelgadoetal94, estebanetal02}. 
\label{tfl_adf}} 
\end{center} 
\end{figure}

The standard temperature fluctuations hypothesis predicts that the ADF 
is related to the difference between {\elect}(CELs) and {\elect}({\hi}) \citep{peimbertcostero69, 
torrespeimbertetal80}. 
Indeed, \citet{liuetal00} found a strong correlation between these two discrepancies from data for PNe that cannot 
be reproduced by the temperature fluctuation paradigm. The results 
of our work seem to indicate that there is also a slight correlation for {\hii} regions, but with a flatter 
slope than 
for PNe (see Figure~\ref{tfl_adf}). 
In Figure~\ref{tfl_adf} we have also included the results obtained for 2 extragalactic {\hii} regions: 
30 Doradus \citep{apeimbert03} and NGC~2363\footnote{For this object the ADF has been obtained from the analysis by 
\citet{estebanetal02}, and temperatures were obtained from \citet{gonzalezdelgadoetal94}. 
We have included these data because they correspond to the same slit position, covering a very 
similar volume 
of nebula.} \citep{gonzalezdelgadoetal94, estebanetal02}. 
In the case of {\hii} regions, contrary to what happens in PNe, the relation between ADF and the temperature 
differences seem to be consistent with the standard temperature fluctuations hypothesis, 
and they are translated into moderate and similar values of the {\ts} parameter, without reaching the extreme 
values found in PNe. 
The fit obtained for our data is: 

\begin{eqnarray} 
\log[{\rm ADF}({\rm O}^{++}) ] & = & (0.184\pm0.022)+(0.609\pm0.208) \\ 
& & \times10^{-4}[T_e({\rm [ O~III]})-T_e({\rm H~I}) ], 
\end{eqnarray} 

\noindent with a correlation coefficient of $r$= 0.575. 

Finally, as we previously pointed out \citep[see][]{garciarojasetal06b}, we do not find the correlation between 
the ADF and the electron density, {\elecd}, obtained by \citet{robertsontessigarnett05} for a PNe sample. 

\section{Objections to the two phase model.
\label{discusiont2}} 

The fact that {\hii} regions as well as PNe are photoionized nebulae does not imply that photoionization is
responsible for all the observed properties of these objects. A common argument against temperature fluctuations in 
{\hii} regions is based on the difficulties of photoionization models to reproduce {\ts} values of the order of those 
estimated for most objects. 
However, photoionization models might not be sufficiently realistic, and/or 
additional energy sources could be necessary to explain the discrepancies between observations and model 
predictions 
\citep[see][and references therein]{viegas02}. Several authors have proposed the existence of 
a plasma component --colder and metal-rich-- (the so-called two-phase model) to explain the dichotomy between 
abundance and temperature determinations 
in PNe, where the standard temperature and/or density fluctuations hypothesis do not seem to explain the available observations 
\citep{liuetal01, tsamisetal04}. 
As we have already commented in \S~\ref{intro}, \citet{tsamispequignot05} have proposed a similar explanation for 
the abundance discrepancy in {\hii} regions although there are not direct observational evidences of the colder 
and metal-rich inclusions they propose\footnotetext{It should be remarked that the hypothesis involved in two phase scenario 
are in the classical temperature fluctuations paradigm are not of the same order, since the two phase scenario actually gives a 
natural explanation for the origin of the temperature fluctuations.}. 
In the following, we are going to present some objections to this model, based on the results obtained in this 
work.

In Table~\ref{adfcomp} we present values of the ADF computed for different ions in Galactic and extragalactic 
{\hii} regions. 
The table includes all the data available up to date, including our own results and others from the literature. 
As we commented before, the ADF for O$^{++}$ is quite similar in all the objects --Galactic or 
extragalactic-- and 
also similar to the ADFs for the other ions: C$^{++}$, O$^+$, and Ne$^{++}$. Comparing the results for {\hii} 
regions of 
different galaxies, 
we find that the ADF seems to be independent of the morphological type, mass, metallicity or even the star 
formation history 
of the host galaxy \citep[see also][]{lopezsanchezetal06}. In the case of the Galactic objects, the similarity of 
the ADFs suggests 
that the process that produces this phenomenon is independent of the conditions and properties of the Galactic 
disk, 
at least along the range of Galactocentric distances covered by our sample.  \citet{lopezsanchezetal06} have 
discussed this fact 
in a paper that presents the ADF values for several zones of NGC 5253, a dwarf Wolf-Rayet galaxy. 

\setcounter{table}{7}
\begin{deluxetable}{cccccccc} 
\tabletypesize{\scriptsize}
\tablecaption{Observed ADFs in Galactic and extragalactic H~{\sc ii} regions.
\label{adfcomp}}
\tablewidth{0pt}
\tablehead{
\colhead{ID} & 
\colhead{{\elecd} (cm$^{-3}$)} & 
\colhead{{\elect}({\foiii}) (K)}  &
\colhead{ADF(O$^{++}$)}  &
\colhead{ADF(O$^{+}$)}  &
\colhead{ADF(C$^{++}$)$^{\rm a}$}  &
\colhead{ADF(Ne$^{++}$)}  &
\colhead{Ref.$^{\rm b}$}  \\ 
\noalign{\smallskip} 
\hline
\noalign{\smallskip} 
\mc{8}{c}{Galactic H~{\sc II} regions}}
\startdata 
M16     & 1120 			& 7650			& 2.8		& \nodata      	& \nodata      	& \nodata      	& 1  	\\
M8     	& 1800 			& 8150			& 2.0		& 1.4	       	& 1.8     	& \nodata    	& 1 	\\
     	& 1750 			& 8050			& 2.0		& \nodata	& 2.1		& \nodata	& 2	\\
M17    	& 470			& 8050			& 2.1		& 4.2:$^{\rm c}$& \nodata      	& \nodata     	& 1 	\\
    	& 860, 520		& 8120, 8210		& 1.8, 2.2	& \nodata	& \nodata	& \nodata	& 3	\\
   	& 600--1500		& 8200			& 2.1		& \nodata	& \nodata	& \nodata	& 4	\\
M20    	& 270			& 7980			& 2.1		& 1.5	       	& \nodata      	& \nodata      	& 1 	\\
NGC~3576& 2300			& 8500			& 1.8		& \nodata      	& \nodata      	& 2.0      	& 1 	\\
	& 1300--2700		& 8850			& 1.8		& \nodata	& \nodata	& \nodata	& 4	\\
Orion	& 7800			& 8320			& 1.4		& 1.6$^{\rm c}$	& 1.9          	& 1.9	     	& 1 	\\
	& 4000, 5700		& 8300, 8350		& 1.3, 1.5	& \nodata      	& 2.5, 2.2	& \nodata      	& 5	\\
NGC~3603& 3400			& 9030			& 1.9		& \nodata      	& \nodata      	& \nodata      	& 1 	\\
S~311	& 310			& 9050			& 1.9		& \nodata      	& \nodata      	& \nodata      	& 1 	\\
\hline
\noalign{\smallskip} 
\multicolumn{8}{c}{Extragalactic H~{\sc II} regions} \\ 
\noalign{\smallskip} 
\hline
\noalign{\smallskip} 
LMC 30 Dor& 370--1800		& 10100			& 2.0--2.7	& \nodata	& \nodata     & \nodata       & 4     	\\
	& 316			& 9950			& 1.6		& \nodata	& \nodata     & \nodata       & 6     	\\
LMC N11B& 80--1700		& 9400			& 4.9--8.2	& \nodata	& \nodata     & \nodata       & 4     	\\
SMC N66 & 50--3700		& 12400			& 2.3		& \nodata	& \nodata     & \nodata       & 4     	\\
NGC~604	& $\le$ 100		& 8150			& 1.6		& \nodata	& \nodata     & \nodata       & 7     	\\
NGC~2363& 360--1200		& 15700			& 2.2		& \nodata	& \nodata     & \nodata       & 7     	\\
NGC~5461& 300			& 8600			& 1.9		& \nodata	& \nodata     & \nodata       & 7     	\\
NGC~5471& 220--1150		& 14100			& 1.6		& \nodata	& \nodata     & \nodata       & 7     	\\
NGC~6822 V& 175			& 11900			& 1.9		& \nodata	& \nodata     & \nodata       & 8     	\\
NGC~5253 A& 580			& 12100			& 1.8		& \nodata	& 2.6	      & \nodata       & 9     	\\
NGC~5253 B& 610			& 12030			& 1.5		& \nodata	& 2.1	      & \nodata       & 9     	\\
NGC~5253 C& 370			& 10810			& 1.9		& \nodata	& 2.1	      & \nodata       & 9     	\\
\enddata
\tablenotetext{a}{CELs abundances for C$^{++}$ were obtained from $IUE$ data of C~{\sc iii}] 
$\lambda\lambda$1906+1909 line. 
\citep[][for M8, the Orion nebula and NGC~5253, respectively]{peimbertetal93b, walteretal92, kobulnickyetal97}.}
\tablenotetext{b}{1) This work; 2) \citet{estebanetal99b}; 3) \citet{estebanetal99a}; 4) \citet{tsamisetal03}; 5) 
\citet{estebanetal98}; 
6) \citet{apeimbert03}; 7) \citet{estebanetal02}; 8) \citet{apeimbertetal05}; 9) \citet{lopezsanchezetal06}.}
\tablenotetext{c}{Dubious value.}
\end{deluxetable}

One of the observational arguments against pure temperature fluctuations in PNe by several authors 
is the agreement between total abundances derived from  
optical and FIR CELs in PNe \citep[see e.g.][]{liuetal01,tsamisetal04}. Due to their low excitation energy, 
E$_{ex}$ $\sim$ 10$^3$ K, the fine structure infrared line emissivities 
have a very weak dependency with {\elect}, therefore they must be insensible to the uncertainties introduced by the 
presence of temperature fluctuations and, in the case of the existence of such fluctuations, they must give values 
of abundance 
similar to that obtained from ORLs. In the case of PNe, this comparison is usually feasible due to the small 
angular size of these objects, both in the optical range
\citep[using scanning techniques to cover the whole nebula, see e.g.][]{liuetal00}, 
and in FIR and UV \citep[see e.g.][]{liuetal00,liuetal01}. 
However, we think that this comparison is not so feasible in the case of extended Galactic {\hii} regions. 
These objects cover much larger sky areas than PNe, 
so large uncertainties may arise due to ionization stratification when comparing ionic abundances derived from 
narrow slit optical spectroscopy and FIR observations obtained with space-borne facilities, that use larger 
apertures. 
In a previous paper \citep{garciarojasetal06} we have showed the difficulties found when comparing our derived 
total 
abundances for three objects of our sample (M17, NGC~3603 and the Orion nebula) with those derived from FIR 
observations 
\citep{simpsonetal95}.

\subsection{Dependence of the ADF on the excitation energy and the critical density.}

\begin{figure}
\begin{center}
\epsscale{1.0}
\plotone{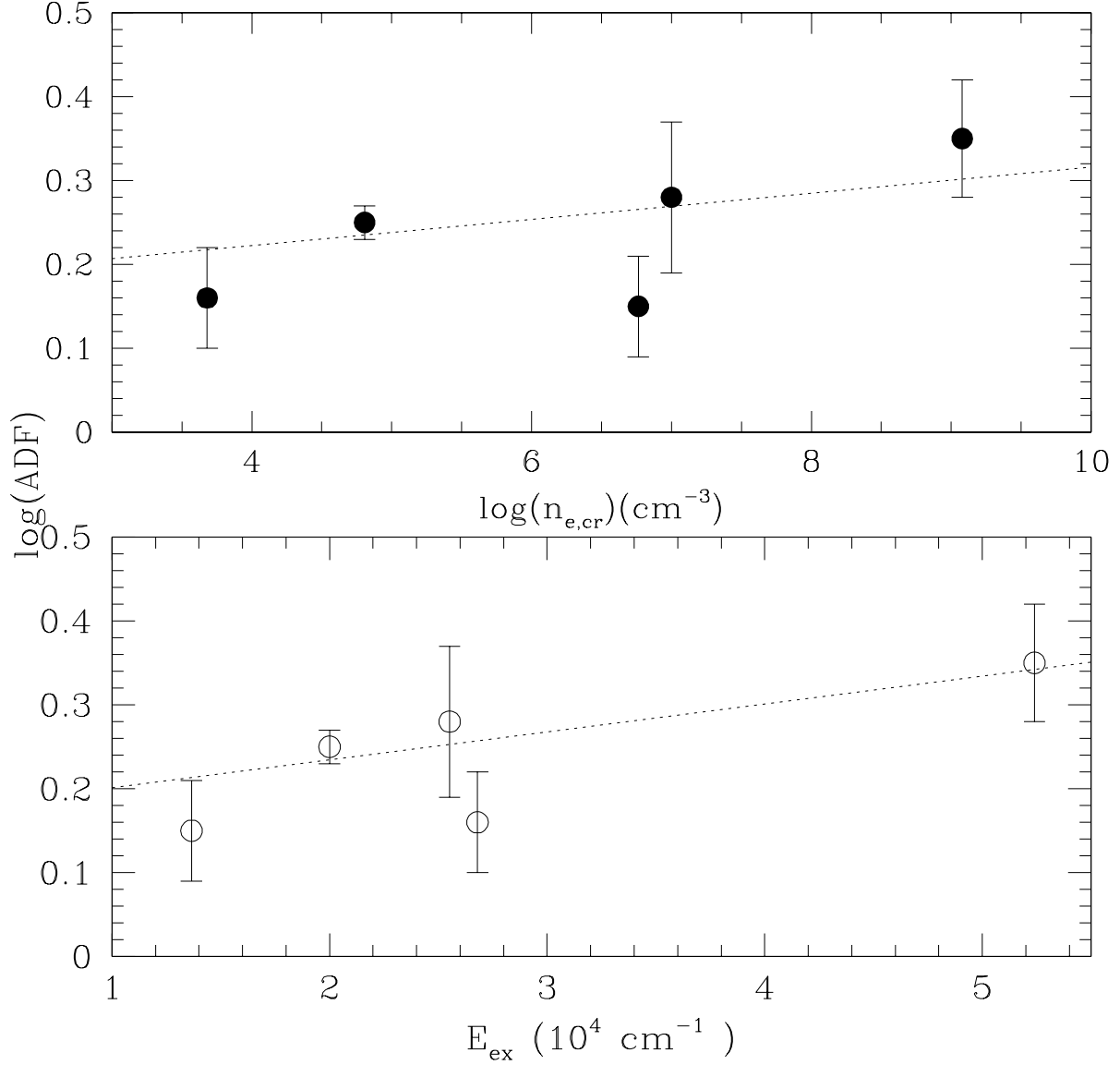} 
\figcaption{ADF (obtained for different ions) $vs.$ excitation energy of the upper level from which each line 
originates (down) and critical density (up). In upper panel, the represented ions are, from left to right: O$^+$ 
({\foii} $\lambda\lambda$ 3726+29), O$^{++}$,  O$^+$ ($\lambda\lambda$ 7319+20+29+30), Ne$^{++}$ and C$^{++}$; in lower 
panel the represented ions are, from left to right: O$^+$ ($\lambda\lambda$ 7319+20+29+30), 
O$^{++}$, Ne$^{++}$, O$^+$ ({\foii} $\lambda\lambda$ 3726+29) and C$^{++}$.
\label{adf_exc_ncr}} 
\end{center} 
\end{figure}

By definition, in the temperature fluctuations formulation, the abundance discrepancy must be related to 
the excitation energy, E$_{ex}$ of the upper level in which the line originates 
\citep[see basic equations in][]{peimbert67}; whereas in the presence of 
density fluctuations the 
abundance discrepancy is maximized if the abundance has been calculated from a CEL coming from a level with low 
critical density 
$n_c$. This last case implies that the ADF and $n_c$ are inversely related. 
\citet{liuetal00, liuetal01}, comparing different abundance determinations in PNe (UV, optical and FIR CELs), 
showed that the 
ADF was not related to E$_{ex}$ nor to $n_c$. We cannot make comparisons between different types of lines 
(UV, optical or FIR) of the same ion, since we do not have observations of the same zones in other spectral range 
than the optical one; however, 
we can compare the ADFs obtained for Galactic and extragalactic {\hii} regions, with E$_{ex}$ 
and $n_c$ of the higher level of the main CELs of each ion. 
In the case of C$^{++}$ the comparison between UV and optical data could be problematic due to the different volumes of nebula covered by the slits. 
For NGC~5253, we are sure that optical and UV observations cover almost the same zones, as was pointed 
out by \citet{lopezsanchezetal06}. In the case of M8 and the Orion nebula, the observations do not cover the same zone, but 
we have compared the ionization correction factor for C$^{++}$ obtained from computing C$^{+}$/H$^{+}$ ratio from $IUE$ observations 
of the UV C~{\sc ii}] $\lambda$2326 line \citep[][for M8 and the Orion nebula, respectively]{peimbertetal93b, walteretal92}  
and by using the ionization correction factor (ICF) of \citet{garnettetal99}. From these data we have obtained ICFs of 2.48 and 2.45 
for M8, and 1.20 and 1.24 for the Orion nebula, suggesting that there are not significant changes in the ionization structure 
of the observed zones in the optical and UV.
In Figure~\ref{adf_exc_ncr} it can be noted that, 
taking into account the uncertainties, a slight correlation ($r$=0.7, $P$=0.188) between the ADF and E$_{ex}$ is apparent. 
There is also an apparent correlation ($r$=0.7, $P$=0.188) with $n_c$, the opposite to that predicted by density fluctuations, 
which suggests that density fluctuations should play a much less important role than temperature fluctuations in 
{\hii} regions. 
These results provide an additional argument that the mechanism that produces the abundance discrepancy in 
{\hii} regions should be different to that 
proposed for PNe and, therefore, we can not make the same considerations when working with PNe and {\hii} regions. 
In Table~\ref{adf_exc}, we show the values of ADF, E$_{ex}$ and $n_c$ adopted for each ion. The values we have adopted are the 
weighted mean of the ADFs computed for each ion in different objects. 
We distinguish between O$^+$ abundances from {\foii} $\lambda\lambda$3726+3729 and $\lambda\lambda$7320+30 because 
they have very different E$_{ex}$ and $n_c$.

\setcounter{table}{8}
\begin{deluxetable}{cccc} 
\tablecaption{ADFs, excitation energy and critical density of the CELs used to compute the abundances.
\label{adf_exc}}
\tablewidth{0pt}
\tablehead{
\colhead{Ion} & 
\colhead{log(ADF)$^{\rm a}$} & 
\colhead{E$_{ex}$ (cm$^{-1}$)} & 
\colhead{$n_c$ (cm$^{-3}$)}}  
\startdata 
C$^{++}$	& 0.35$\pm$0.07	& 52400	 & 1.2$\times$10$^9$  \\
Ne$^{++}$	& 0.28$\pm$0.09	& 25520	 & 1.0$\times$10$^7$  \\
O$^{++}$	& 0.25$\pm$0.02	& 20000	 & 6.4$\times$10$^4$  \\
O$^{+}$$^{\rm b}$& 0.16$\pm$0.06& 26800	 & 4.8$\times$10$^3$  \\
O$^{+}$$^{\rm c}$& 0.15$\pm$0.06& 13660	 & 5.8$\times$10$^6$  \\
\enddata
\tablenotetext{a}{Weighted mean for all Galactic and extragalactic H~{\sc ii} regions.}
\tablenotetext{b}{From [O~{\sc ii}] $\lambda\lambda$ 3726+29.}
\tablenotetext{c}{From [O~{\sc ii}] $\lambda\lambda$ 7319+20+29+30.}
\end{deluxetable}

\subsection{ {\oii} ORLs temperature.
\label{te_rls}} 

The two phase model predicts that {\elect}(ORLs) $\le$ {\elect}({\hei}) $\le$ {\elect}({\hi}) $\le$ {\elect}(CELs) 
\citep{liu03}, 
with the difference between any of these two temperatures being proportional to the ADF (see e.g. 
Figure~\ref{tfl_adf}). 

\citet{wessonetal03} made use --for the first time-- of the temperature sensitive ratio 
$I$($\lambda$4089.29)/$I$($\lambda$4649.14) to determine {\elect} of the ionized gas in which {\oii} lines arise in 
PNe.  
These authors found very low {\elect} in two H-deficient knots in the PN Abell 30. Later, several authors have 
found 
similar results in other PNe \citep[e.g.][]{tsamisetal04, yliuetal04b, wessonetal05}. The intensities of the ORLs 
that 
originate from states with different orbital angular momentum present different dependencies with {\elect}. 
For example, comparing the intensity of one 3$d$--4$f$ transition with respect to a 3$s$--3$p$ transition, is 
possible to compute 
the electron temperature. However, this method has its difficulties: first, the dependency of the intensity ratio 
with the 
temperature is weak, so these lines must be measured very accurately; second, the relative 
intensities of the {\oii} lines can be affected by departures of the local thermodynamic equilibrium (LTE) 
in the fundamental recombination level of the O$^{++}$ ion, $^3$P. \citet{tsamisetal04} consider the 
ratio of the 3$d$--4$f$ transition of {\oii} $\lambda$4089.29 and the 3$p$--3$s$ transition {\oii} $\lambda$4649.14 
of multiplet 1 
because these lines arise from states of high total angular momentum J, $^3$P$_2$ and, therefore they must be 
affected in a similar way by this effect. 

\begin{figure*}
\begin{center}
\epsscale{0.48}
\plotone{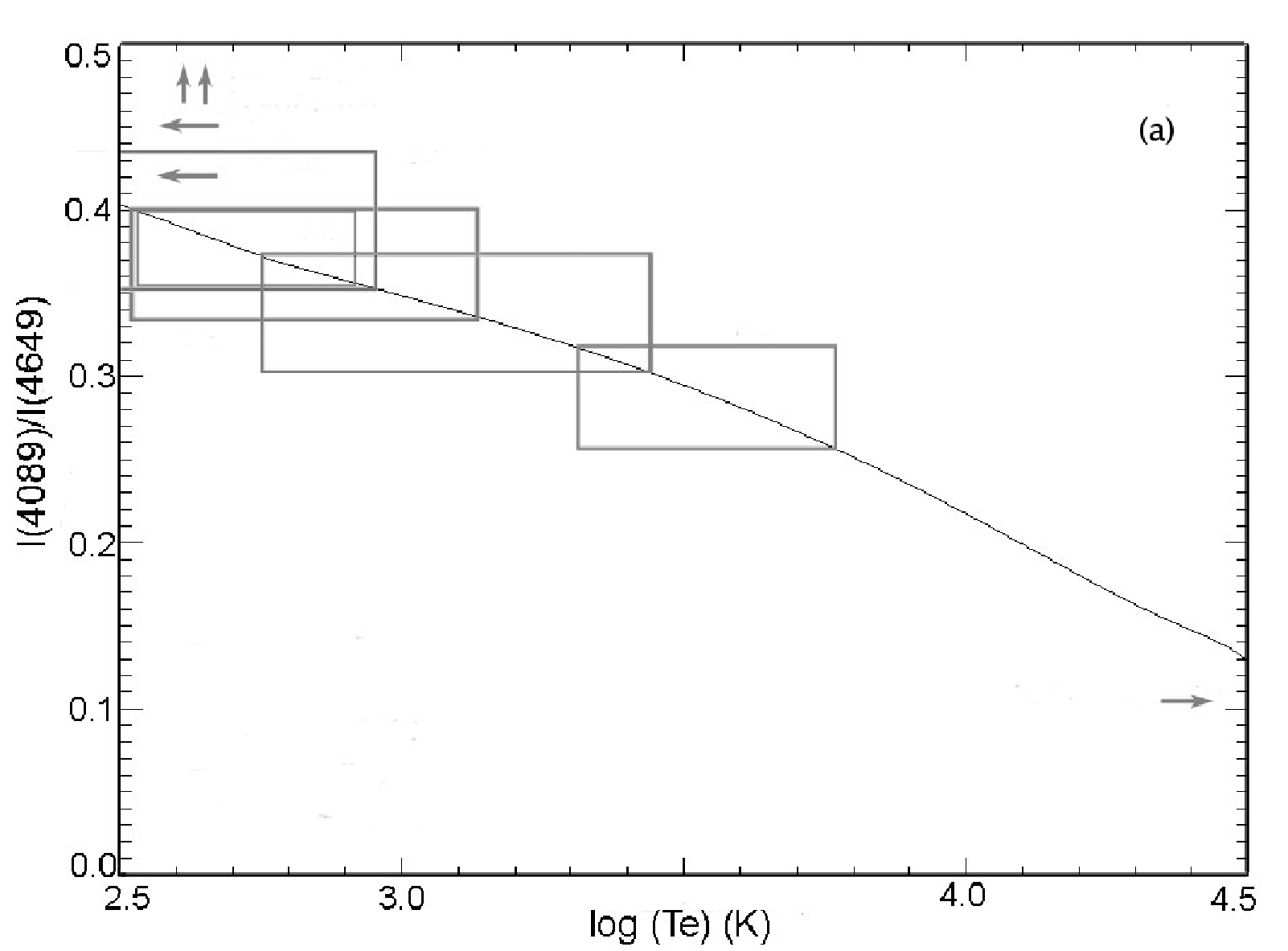}
\plotone{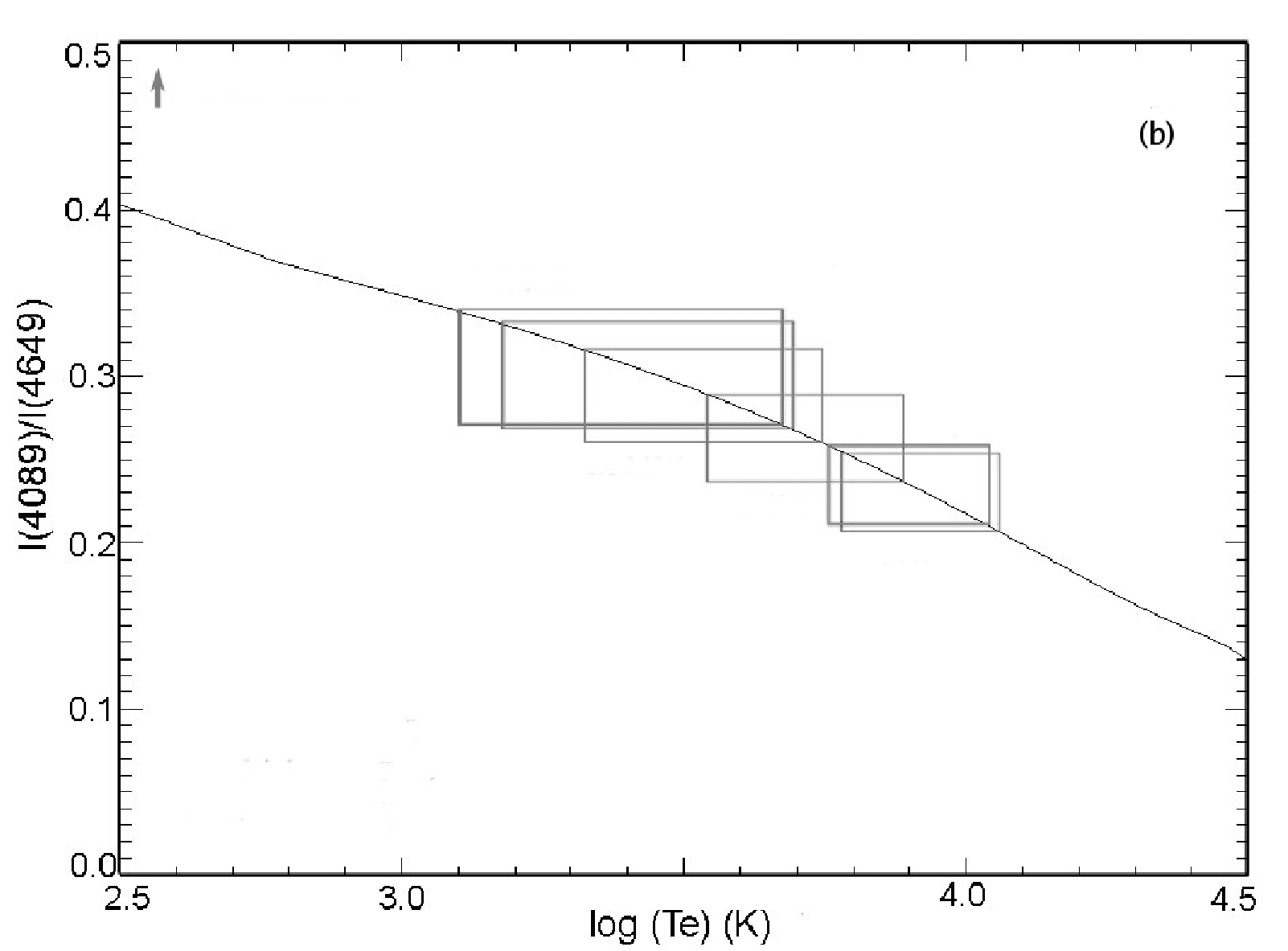} 
\plotone{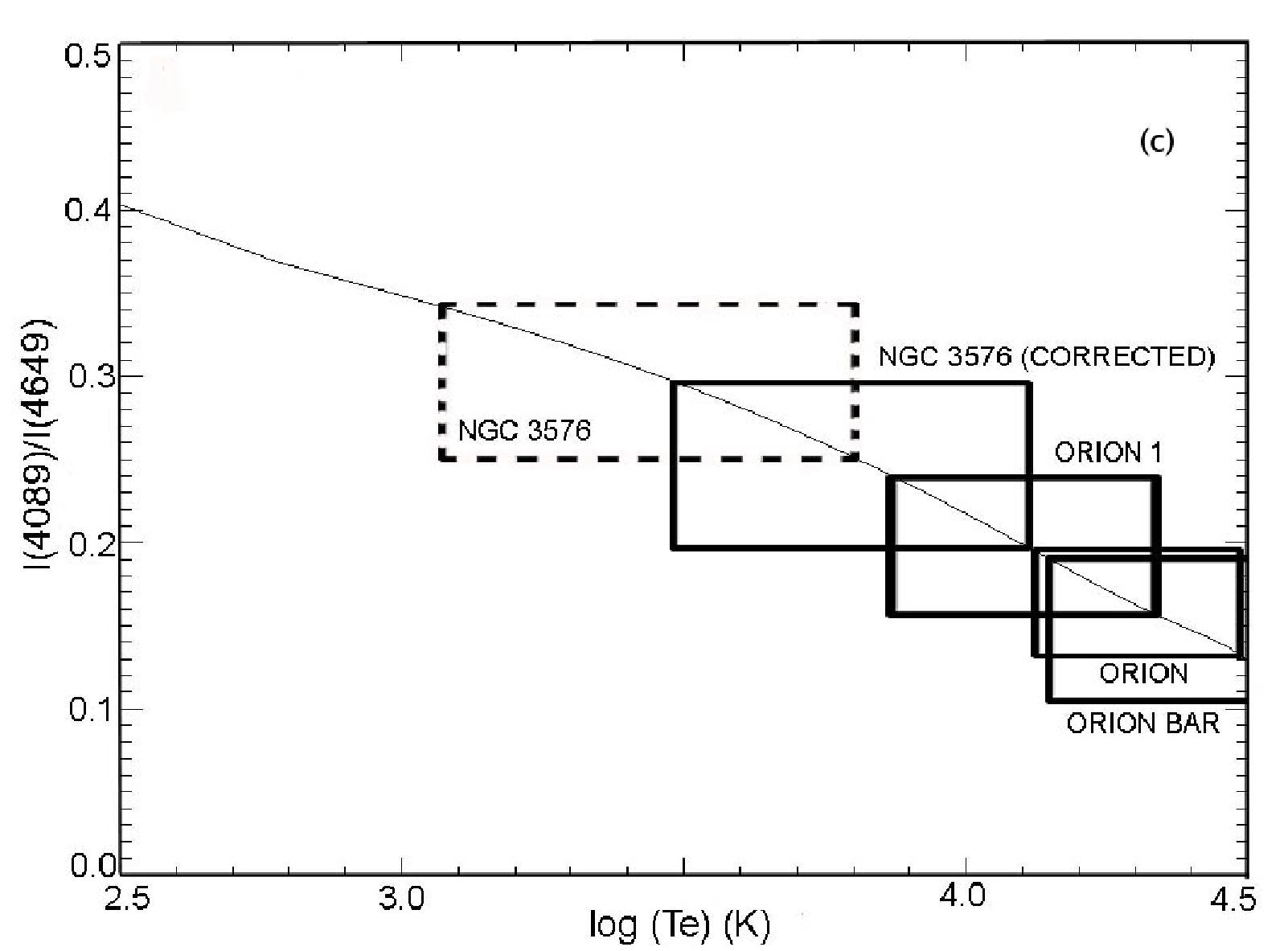} 
\plotone{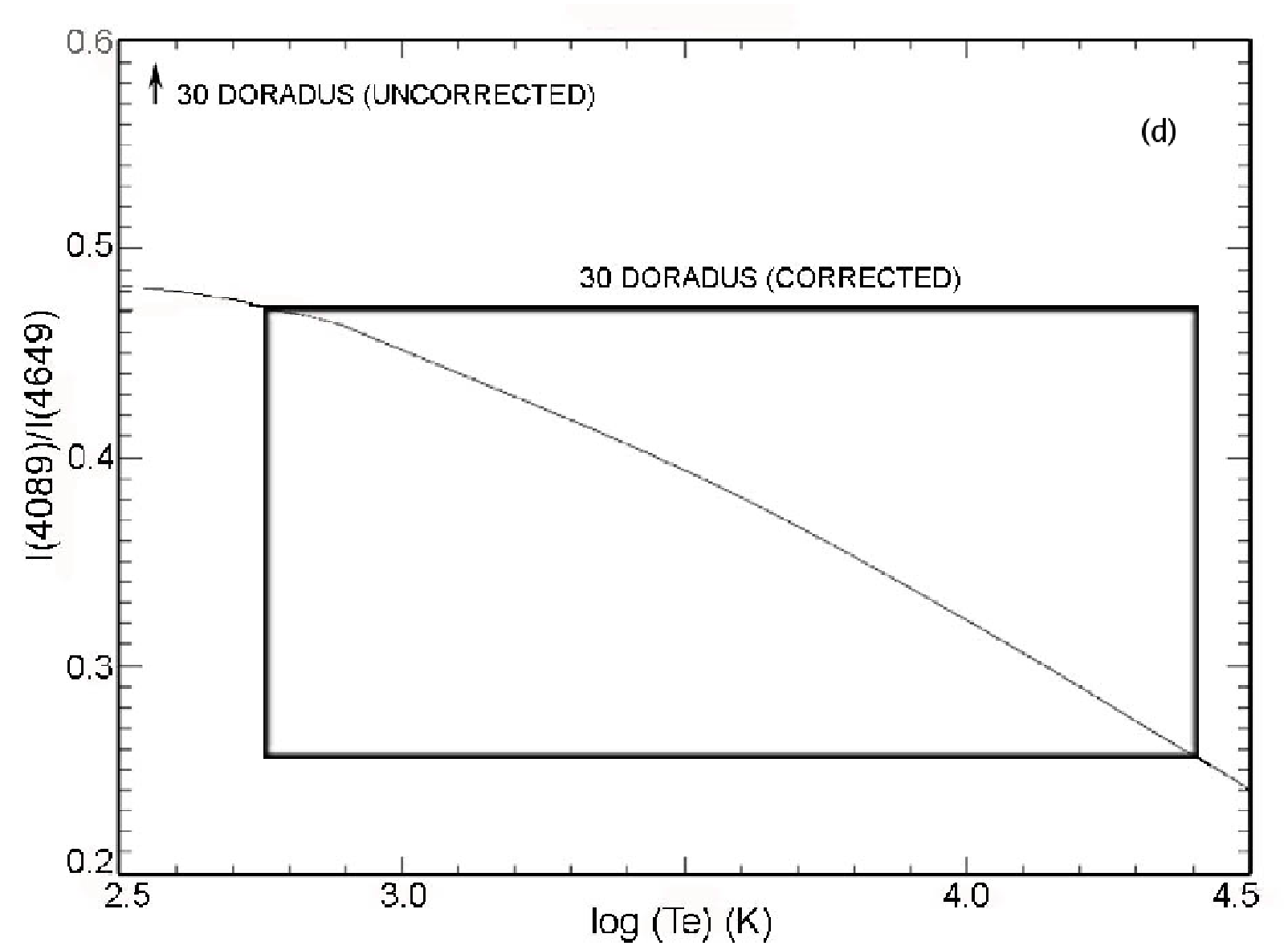} 
\figcaption{O~{\sc ii} $\lambda$4089.29/$\lambda$4649.14 line ratio as a function of the 
electron temperature. Solid line in a), b)  and c) panels is the theoretical ratio for {\elecd} = 10$^4$ cm$^{-3}$. 
The data are represented as error boxes (see text for the references). 
In panel a) we have represented a sample of PNe with ADF $>$ 5.0; in panel b), we have 
represented a sample of PNe  with moderate ADFs, similar to those found in our sample of H~{\sc ii} regions; in the 
panel c), we show the two H~{\sc ii} regions for which it has been possible to measure O~{\sc ii} $\lambda$4089.29 3d--4f 
transition: NGC~3576 and the Orion Nebula and two additional slit positions in the Orion nebula (see text) and, 
finally, in panel d), we show the data for 30 Doradus \citep{apeimbert03}. 
In this last diagram we have also included the theoretical behavior of the line ratio 
with respect to electron temperature for {\elecd} = 250 cm$^{-3}$ (P.J. Storey, private communication) in order to 
perform a more appropiate comparison in the case of 30 Doradus, which shows {\elecd} $\sim$ 300 cm$^{-3}$.
For NGC~3576 and 30 Doradus, solid line boxes indicate the values found after correction by the presence of 
charge transfer features in the CCD (see text), the dashed box indicates the position of NGC~3576 before the 
correction. The horizontal arrows indicate the value of the line ratio for some objects which corresponding 
temperature is outside the scale, whereas the vertical arrows indicate that the line ratio is outside the 
represented scale. 
\label{temp_rls}} 
\end{center} 
\end{figure*}

\begin{figure}
\begin{center}
\epsscale{1.0}
\plotone{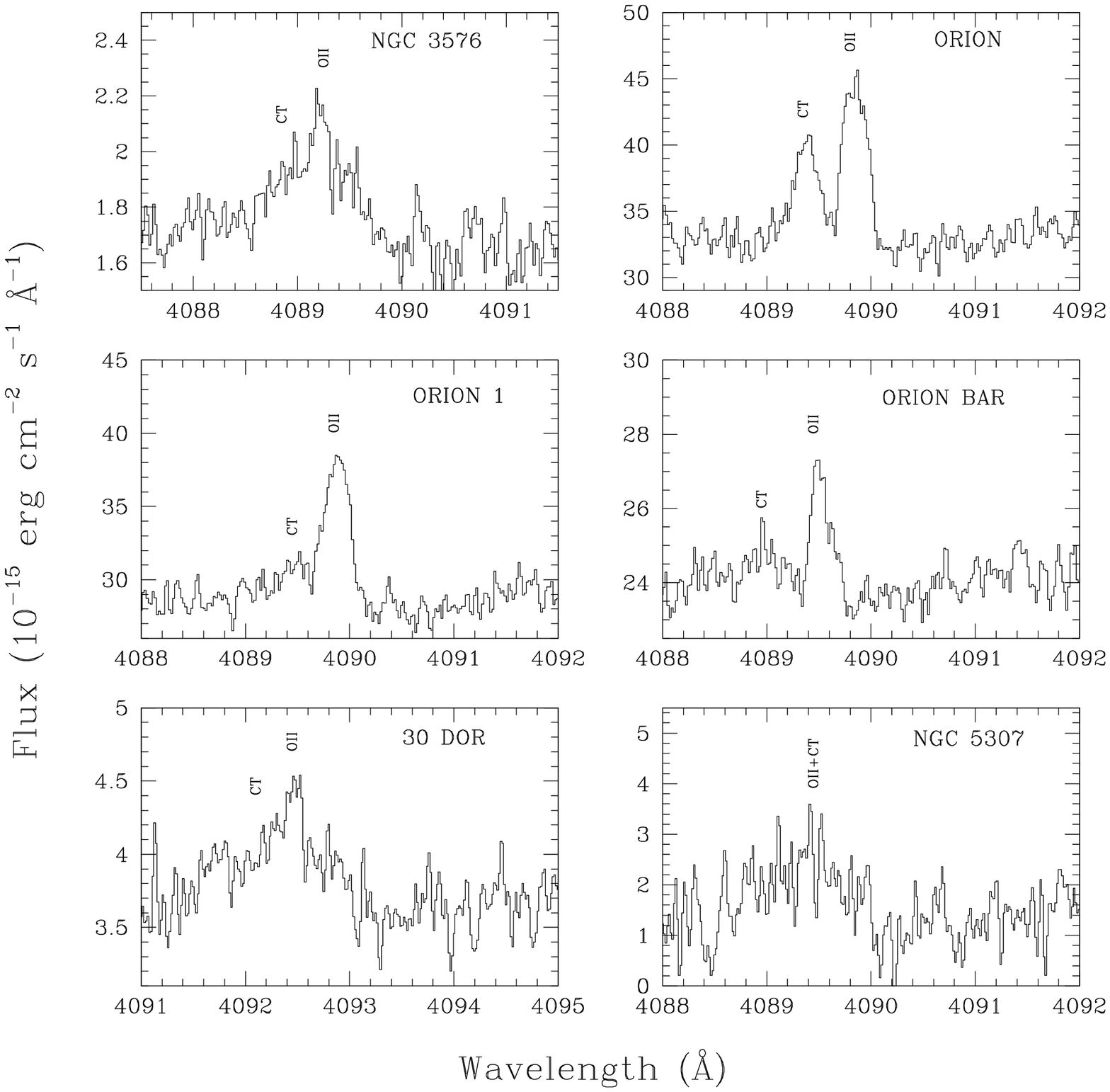} 
\figcaption{ O~{\sc ii} $\lambda$4089.19 line in 5 spectra of H~{\sc ii} regions 
and in PNe NGC~5307. It can be noticed that this line is severely blended with another feature produced by charge 
transfer along the CCD (CT) in NGC~3576 and 30 Doradus, whereas in NGC~5307 the signal to noise is too low. 
The Orion nebula is the only H~{\sc ii} region in which this line has been well measured. 
The calibrated spectra of 30 Doradus and NGC~5307 have been kindly provided by Antonio Peimbert. 
\label{linea4089}} 
\end{center} 
\end{figure}

Determination of electron temperature from {\oii} ORLs in {\hii} regions is a non trivial problem because in these 
objects ORLs are usually much weaker than in PNe. In Figure~\ref{temp_rls} we have represented the change of 
$I$($\lambda$4089.29)/$I$($\lambda$4649.14) ratio with {\elect} for {\elect}=10$^4$ cm $^{-3}$, and we have 
compared 
that function with 
the ratios obtained in NGC~3576 and the Orion nebula, the only objects of our sample where {\oii} $\lambda$4089.29 
line has been measured. Additionally, we have represented the ratio measured in the giant {\hii} region 30 Doradus 
in the 
Large Magellanic Cloud (LMC) \citep{apeimbert03} and in several PNe taken from the literature: M~1-42 and M~2-36 
\citep{liuetal01}, \citet{tsamisetal04} sample, PNe with ADF $>$ 4 of \citet{wessonetal05} sample, 
NGC~5307 \citep{ruizetal03} and NGC~5315 \citep{peimbertetal04}. This last PN is also included in the PNe 
sample of \citet{tsamisetal04} and we have considered it with the purpose of comparing the effect that 
high spectral resolution observations introduce in the determination of {\elect}({\oii}). 
NGC~5315 is appropiate because NLTE effects are small due to their high {\elecd} ($\sim$10$^4$ cm$^{-3}$) and 
also presents an ADF which is similar to that measured in {\hii} regions. 
Within the sample of PNe included in panel a) of Figure~\ref{temp_rls} there are objects with very high ADF 
(between 5.0 and 
22) and in panel b) of Figure~\ref{temp_rls} there are other PNe that show values similar to that obtained in 
{\hii} regions. NGC~5307 and NGC~5315 show a moderate 
ADF (1.9 and 1.7 
respectively), and have been studied from high resolution spectra (R$\sim$8800), which avoids the overlapping of 
{\oii} multiplet 1 lines (in the other PNe, {\oii} $\lambda$4649.14 has been measured deblending the line using a 
multiple Gaussian fit) and of {\oii} $\lambda$4089.29 line with other spectral features\footnote{In some of the PNe 
of the 
Tsamis et al. sample, the contribution of Si~{\sc iv} $\lambda$4088.85 line to the intensity of O~{\sc ii} 
$\lambda$4089.29 
line was corrected using the measured intensity of Si~{\sc iv} $\lambda$4116.10 line and adopting the theoretical 
Si~{\sc iv} $\lambda$4088.85/Si~{\sc iv} $\lambda$4116.10 flux ratio (2:1).}. 
As we can see in panel a) of Figure~\ref{temp_rls}, PNe with high ADFs show, in general, very low 
{\elect}({\oii}), 
which is consistent with the two phase model predictions. On the other hand, PNe with moderate ADFs (panel b) 
of Figure~\ref{temp_rls}) 
present temperatures that, in general, are also below {\elect}(CELs) and {\elect}({\hi}), except in the cases 
of NGC~5882 \citep{tsamisetal04} and NGC~5315 \citep{peimbertetal04}, where {\elect}({\oii}) is consistent with 
the derived {\elect} 
from CELs. On the other hand, {\elect}({\oii}) obtained for NGC~5307 is extremely low, which can be due to 
uncertainties in the measurement of the intensity of {\oii} $\lambda$4089.29 line\footnote{The NGC~5307 spectrum 
was obtained in the same observation run as those of NGC~3576 and the Orion nebula. In the analyzed spectra of these 
observations, 
O~{\sc ii} $\lambda$4089.29 line is closely blended with a spectral feature produced by charge transfer effects in 
the CCD due to the presence of a very bright line in a different order; this feature is clearly separated from the 
O~{\sc ii} 
$\lambda$4089.29 line in the Orion nebula, but not in NGC~3576 nor in 30 Doradus 
(see Figure~\ref{linea4089}). An alternative form to correct 
it, would be to adopt the theoretical ratio with other line of the same multiplet but unfortunately we do not know 
the effect of NLTE to the 
relative intensity of the lines of the multiplet.} (see Figure~\ref{linea4089}). In the case of the Galactic 
{\hii} regions, 
{\elect}({\oii}) obtained for the Orion nebula is somewhat higher to that obtained from CELs; for NGC~3576, 
{\elect}({\oii}) is smaller than that 
obtained from CELs, but {\oii} $\lambda$4089.29 line is affected by charge transfer effects in the CCD (see 
Figure~\ref{linea4089}). 
Similarly, $I$(4089)/$I$(4649)=0.825 ratio found for 30 Doradus \citep{apeimbert03} is extremely high, 
due to a significant contribution of charge transfer to the intensity of the line (see Figure~\ref{linea4089}). 
For these two regions we have tried to separate the line from the parasite feature, but it was not easy because of 
the low signal-to-noise of the data in the spectral range of interest. As a first step we have identified the line 
responsible of the 
charge transfer effect in the CCD and then we have measured the flux of all the features 
produced by charge transfer in the CCD (hereinafter CCT) in all the orders; finally, we have fitted the flux 
decrement through the 
different orders in order to 
estimate the contribution of the CCT to the {\oii} $\lambda$4089.29 line. 
Unfortunately, uncertainties 
in the observed fluxes were high, so we can only make a rough estimation to the corrected line flux. For NGC~3576, 
we have 
estimated that CCT is about a 16\% of the observed flux of the {\oii} $\lambda$4089.29 line. 
In Figure~\ref{temp_rls}c we have represented the new corrected value for NGC~3576, adopting the same uncertainty 
in the line 
flux that before correction. 
In the case of 30 Doradus, we have estimated that the contribution to the observed flux of 
{\oii} $\lambda$4089.29 line is between 30\% and 60\%.  
We have represented an intermediate case (correction of 45\%) in Figure~\ref{temp_rls}d, assuming that the 
uncertainty in the 
$I$(4089)/$I$(4649) ratio of 30 Dor is about 30\% or even greater. 
As it can be seen in Figure~\ref{temp_rls}c, applying the estimated corrections, 
the {\elect}({\oii}) obtained for NGC~3576 is now consistent with {\elect} derived from CELs. For 30 Doradus, 
taking into 
account that the {\elecd} derived by \citet{apeimbert03} was $\sim$300 cm$^{-3}$, we have represented 
the theoretical $I$($\lambda$4089.29)/$I$($\lambda$4649.14) ratio for {\elecd}= 250 cm$^{-3}$ (P.J. Storey, private 
communication) 
which is quantitatively different from that derived for {\elecd}=10$^4$ cm$^{-3}$; unfortunately, in this case, 
the results are not as clear as in the case of NGC~3576: the large uncertainties in the adopted flux of the 
{\oii} $\lambda$4089.29 line makes the {\oii} $I$(4089)/$I$(4649) ratio unreliable to compute {\elect} because 
uncertainties 
are compatible with variations of {\elect} from 600 K to 25000 K (see Figure~\ref{temp_rls}d). 
In order to have additional {\hii} region data, we have represented the value of the intensity ratio for 
non-published UVES datasets for two slit positions of the Orion nebula: 
the position labeled as ``Orion 1'' coincides with one of the two zones studied by \citet{estebanetal98}, 
and one slit position located on the Orion bar (24$\arcsec$ N and 12$\arcsec$ W of $\theta^2$Ori A). 
From the position of the error boxes for these two regions in Figure~\ref{temp_rls}c, it can be seen 
that the values of {\elect} are similar to that obtained for the Orion 2 slit position. 
Considering {\elect}({\foiii}) calculated for these two regions and the relatively high densities that have been 
computed in both regions 
--{\elecd}({\fcliii}) = 7900$\pm$1300 and 5300$\pm$800 cm$^{-3}$ for Orion 1 and Bar, respectively, which minimizes 
NLTE effects in the intensity of the lines--, we can affirm that {\elect}({\oii}) measured in different positions 
from the Orion nebula are similar or even somewhat larger than that measured from CELs. On the other hand, it is true that
in the presence of pure temperature fluctuations, {\elect}({\oii}) should be somewhat lower than that measured from CELs but, 
taking into account that measured {\ts}'s in {\hii} regions are moderate, differences as large as those predicted by the two-phase 
model are not expected.

Unfortunately, the available data for {\hii} regions are not enough to draw definitive general conclusions; 
however our analysis allows us to enumerate a series of problems that we have found and to reach some 
partial conclusions; a) first, it is necessary to develop atomic models that consider departures from LS coupling 
in NLTE, with the purpose of 
studying how this effect affects each line; b) it is necessary to carry out very deep high resolution spectrophotometry 
in {\hii} regions to extend the catalogue of this type of objects with measured {\elect}({\oii}); c) the few 
available data for {\hii} regions behave differently from those of most PNe and do not follow the predictions of the 
two phase model as given in \citet{tsamispequignot05} for 30 Doradus.

Finally, a similar diagram has been constructed using {\cii} RLs ratios: $I$($\lambda$9903)/$I$($\lambda$4267) and 
$I$($\lambda$6462)/$I$($\lambda$4267) 
--that are not affected by NLTE effects-- but these ratios are much less temperature sensitive than 
{\oii} $\lambda$4089.29/$\lambda$4649.14, and variations between 10$^3$ to 
15$\times$10$^3$ K are of the order of the observed error, so these ratios are not reliable to determine accurately 
electron temperatures.

\section{Conclusions.
\label{conclu}}

We have derived the abundance discrepancy factor, ADF, defined as: O$^{++}$/H$^+$(ORLs)/O$^{++}$/H$^+$(CELs) in eight 
Galactic {\hii} regions, finding similar values for all of them. Furthermore, these values are similar with the ADF 
determinations in extragalactic {\hii} regions available in the literature. This is a fairly different behavior 
from the one shown by the ADF in PNe.

We have found that --within the uncertainties-- the ADF is not related to O/H, O$^{++}$/H$^+$, nor to the ionization degree. 
As well, it is not related either to the assumed 
{\elect}(High), so systematic effects in the abundance determination from CELs can be discarded. Also, the ADF is 
not related to the {\elect}(Low)/{\elect}(High) ratio, suggesting that {\elect} natural gradients in {\hii} regions do 
not produce the abundance discrepancies. 
 
In the case of {\hii} regions, contrary to the case of PNe, the ADF seems to be metallicity independent in the studied 
range of O abundances. Also, there is no 
evidence that the ADF is related to the effective temperature of ionizing stars --as should be implied by 
photoelectric heating due to  the presence of dust near the stars--. 
We have explored the behavior of the ADF with respect to the FWHM of the spectral lines finding no correlation 
although the result is not conclusive due to the paucity of objects with lines of large FWHM.
 
We have found that the ADF seems to be slightly related to the excitation energy, $E_{ex}$ and to the critical density, $n_c$ 
(in contrast to that observed in PNe), 
a behavior that is not predicted by the two phase model initially proposed for PNe and recently extended to {\hii} 
regions \citep{tsamispequignot05}, but consistent --at least in the first aspect-- with the predictions of the standard 
temperature fluctuations paradigm. 
On the other hand, electron temperatures derived from ORLs are consistent 
with those derived from CELs, within the uncertainties, also contradicting the large differences predicted by the two phase model. 
All these results suggest that the physical mechanism that produces the abundance discrepancy in PNe and {\hii} 
regions might be different. Our results do not permit to discern what is the mechanism or 
natural phenomenon that underlies the abundance discrepancy problem, but seem to be more consistent with the predictions of the 
pure temperature fluctuations paradigm than with those of the two phase model. Therefore, it seems necessary to explore further 
other alternatives for the origin of the temperature fluctuations and the abundance discrepancy in {\hii} regions.
 
Obtaining deep spectra of extragalactic {\hii} regions is fundamental to check the different models proposed to 
explain the abundance discrepancy, permitting to extend the parameters space (metallicity, ionization budget, 
kinematics, morphological complexity) of the objects. 

Finally, we want to stress again that, although the presence and survival of temperature fluctuations in gaseous 
nebulae is a controversial fact, it is clear that the behavior in {\hii} regions and PNe should be explored in 
independent ways, due to 
their different origin, evolution time scales and physical processes that could affect both type of objetcs.

JGR and CE would like to thank M. Peimbert, A. Peimbert, M. Rodr\'{\i}guez, A. Mampaso, V. Luridiana, E. P\'erez 
and P.J. Storey for fruitful discussions and comments. We want also to thank all the organizers and participants to the 
workshop on Deep Spectroscopy and Modeling of Gaseous Nebulae (held at Xiang Shan, Beijing on Apr 16-18 2007) for fruitful 
discussions and 
advice related to this work. We would also thank the referee, G. Stasi\'nska, for her carefully reading of the paper and 
for her comments, that have increased significantly the quality of the paper. 
This work has been funded by the Spanish Ministerio de Ciencia y Tecnolog\'{\i}a (MCyT) under project 
AYA2004-07466.


\begin{thebibliography}{65}
\expandafter\ifx\csname natexlab\endcsname\relax\def\natexlab#1{#1}\fi

\bibitem[{{Aller} \& {Menzel}(1945)}]{allermenzel45}
{Aller}, L.~H. \& {Menzel}, D.~H. 1945, ApJ, 102, 239

\bibitem[{{Esteban} {et~al.}(2004){Esteban}, {Peimbert}, {Garc{\'{\i}}a-Rojas},
  {Ruiz}, {Peimbert}, \& {Rodr{\'{\i}}guez}}]{estebanetal04}
{Esteban}, C., {Peimbert}, M., {Garc{\'{\i}}a-Rojas}, J., {Ruiz}, M.~T.,
  {Peimbert}, A., \& {Rodr{\'{\i}}guez}, M. 2004, MNRAS, 355, 229

\bibitem[{{Esteban} {et~al.}(1998){Esteban}, {Peimbert}, {Torres-Peimbert}, \&
  {Escalante}}]{estebanetal98}
{Esteban}, C., {Peimbert}, M., {Torres-Peimbert}, S., \& {Escalante}, V. 1998,
  MNRAS, 295, 401

\bibitem[{{Esteban} {et~al.}(1999{\natexlab{a}}){Esteban}, {Peimbert},
  {Torres-Peimbert}, \& {Garc\'{\i}a-Rojas}}]{estebanetal99a}
{Esteban}, C., {Peimbert}, M., {Torres-Peimbert}, S., \& {Garc\'{\i}a-Rojas},
  J. 1999{\natexlab{a}}, Rev. Mexicana Astron. Astrofis., 35, 65

\bibitem[{{Esteban} {et~al.}(1999{\natexlab{b}}){Esteban}, {Peimbert},
  {Torres-Peimbert}, {Garc\'{\i}a-Rojas}, \& {Rodr\'{\i}guez}}]{estebanetal99b}
{Esteban}, C., {Peimbert}, M., {Torres-Peimbert}, S., {Garc\'{\i}a-Rojas}, J.,
  \& {Rodr\'{\i}guez}, M. 1999{\natexlab{b}}, ApJS, 120, 113

\bibitem[{{Esteban} {et~al.}(2002){Esteban}, {Peimbert}, {Torres-Peimbert}, \&
  {Rodr{\'{\i}}guez}}]{estebanetal02}
{Esteban}, C., {Peimbert}, M., {Torres-Peimbert}, S., \& {Rodr{\'{\i}}guez}, M.
  2002, ApJ, 581, 241

\bibitem[{{Garc{\'{\i}}a-Rojas} {et~al.}(2005){Garc{\'{\i}}a-Rojas}, {Esteban},
  {Peimbert}, {Peimbert}, {Rodr{\'{\i}}guez}, \& {Ruiz}}]{garciarojasetal05}
{Garc{\'{\i}}a-Rojas}, J., {Esteban}, C., {Peimbert}, A., {Peimbert}, M.,
  {Rodr{\'{\i}}guez}, M., \& {Ruiz}, M.~T. 2005, MNRAS, 362, 301

\bibitem[{{Garc{\'{\i}}a-Rojas} {et~al.}(2007){Garc{\'{\i}}a-Rojas}, {Esteban},
  {Peimbert}, {Rodr{\'{\i}}guez}, {Peimbert}, \& {Ruiz}}]{garciarojasetal06b}
{Garc{\'{\i}}a-Rojas}, J., {Esteban}, C., {Peimbert}, A., {Rodr{\'{\i}}guez},
  M., {Peimbert}, M., \& {Ruiz}, M.~T. 2007, Rev. Mexicana Astron. Astrofis.,
  43, 3

\bibitem[{{Garc{\'{\i}}a-Rojas} {et~al.}(2006){Garc{\'{\i}}a-Rojas}, {Esteban},
  {Peimbert}, {Costado}, {Rodr{\'{\i}}guez}, {Peimbert}, \&
  {Ruiz}}]{garciarojasetal06}
{Garc{\'{\i}}a-Rojas}, J., {Esteban}, C., {Peimbert}, M., {Costado}, M.~T.,
  {Rodr{\'{\i}}guez}, M., {Peimbert}, A., \& {Ruiz}, M.~T. 2006, MNRAS, 368,
  253

\bibitem[{{Garc{\'{\i}}a-Rojas} {et~al.}(2004){Garc{\'{\i}}a-Rojas}, {Esteban},
  {Peimbert}, {Rodr{\'{\i}}guez}, {Ruiz}, \& {Peimbert}}]{garciarojasetal04}
{Garc{\'{\i}}a-Rojas}, J., {Esteban}, C., {Peimbert}, M., {Rodr{\'{\i}}guez},
  M., {Ruiz}, M.~T., \& {Peimbert}, A. 2004, ApJS, 153, 501

\bibitem[{{Garnett}(1992)}]{garnett92}
{Garnett}, D.~R. 1992, AJ, 103, 1330

\bibitem[{{Garnett} \& {Dinerstein}(2001)}]{garnettdinerstein01}
{Garnett}, D.~R. \& {Dinerstein}, H.~L. 2001, ApJ, 558, 145

\bibitem[{{Garnett} {et~al.}(1999){Garnett}, {Shields}, {Peimbert},
  {Torres-Peimbert}, {Skillman}, {Dufour}, {Terlevich}, \&
  {Terlevich}}]{garnettetal99}
{Garnett}, D.~R., {Shields}, G.~A., {Peimbert}, M., {Torres-Peimbert}, S.,
  {Skillman}, E.~D., {Dufour}, R.~J., {Terlevich}, E., \& {Terlevich}, R.~J.
  1999, ApJ, 513, 168

\bibitem[{{Gonz\'alez-Delgado} {et~al.}(1994){Gonz\'alez-Delgado}, {P\'erez},
  {Tenorio-Tagle}, {V\'{\i}lchez}, {Terlevich}, {Terlevich}, {Telles},
  {Rodr\'{\i}guez-Espinosa}, {Mas-Hesse}, {Garc\'{\i}a-Vargas}, {D\'{\i}az},
  {Cepa}, \& {Casta\~neda}}]{gonzalezdelgadoetal94}
{Gonz\'alez-Delgado}, R.~M., {P\'erez}, E., {Tenorio-Tagle}, G.,
  {V\'{\i}lchez}, J.~M., {Terlevich}, E., {Terlevich}, R., {Telles}, E.,
  {Rodr\'{\i}guez-Espinosa}, J.~M., {Mas-Hesse}, M., {Garc\'{\i}a-Vargas},
  M.~L., {D\'{\i}az}, A.~I., {Cepa}, J., \& {Casta\~neda}, H. 1994, ApJ, 437,
  239

\bibitem[{{Guseva} {et~al.}(2007){Guseva}, {Izotov}, {Papaderos}, \&
  {Fricke}}]{gusevaetal07}
{Guseva}, N.~G., {Izotov}, Y.~I., {Papaderos}, P., \& {Fricke}, K.~J. 2007,
  A\&A, in press, astro-ph/0701032

\bibitem[{{Guseva} {et~al.}(2006){Guseva}, {Izotov}, \& {Thuan}}]{gusevaetal06}
{Guseva}, N.~G., {Izotov}, Y.~I., \& {Thuan}, T.~X. 2006, ApJ, 644, 890

\bibitem[{{H\"agele} {et~al.}(2006){H\"agele}, {P\'erez-Montero}, {D\'{\i}az},
  {Terlevich}, \& {Terlevich}}]{hageleetal06}
{H\"agele}, G.~F., {P\'erez-Montero}, E., {D\'{\i}az}, A.~I., {Terlevich}, E.,
  \& {Terlevich}, R. 2006, MNRAS, 372, 293

\bibitem[{{Harrington} {et~al.}(1980){Harrington}, {Lutz}, {Seaton}, \&
  {Stickland}}]{harringtonetal80}
{Harrington}, J.~P., {Lutz}, J.~H., {Seaton}, M.~J., \& {Stickland}, D.~J.
  1980, MNRAS, 191, 13

\bibitem[{{Kaler}(1986)}]{kaler86}
{Kaler}, J.~B. 1986, ApJ, 308, 337

\bibitem[{{Kobulnicky} {et~al.}(1997){Kobulnicky}, {Skillman}, {Roy}, {Walsh},
  \& {Rosa}}]{kobulnickyetal97}
{Kobulnicky}, H.~A., {Skillman}, E.~D., {Roy}, J.-R., {Walsh}, J.~R., \&
  {Rosa}, M.~R. 1997, ApJ, 477, 679

\bibitem[{{Liu}(2003)}]{liu03}
{Liu}, X.-W. 2003, in IAU Symp. 209, Planetary Nebulae: Their Evolution and
  Role in the Universe (San Francisco: ASP), ed. S.~{Kwok}, M.~{Dopita}, \&
  R.~{Sutherland}, 339

\bibitem[{{Liu}(2006)}]{liu06}
{Liu}, X.-W. 2006, in IAU Symp. 234, Planetary Nebulae in our Galaxy and Beyond
  (San Francisco: ASP), ed. M.~J. {Barlow} \& R.~H. {M\'endez}, 219

\bibitem[{{Liu} {et~al.}(2006){Liu}, {Barlow}, {Zhang}, {Bastin}, \&
  {Storey}}]{liuetal06}
{Liu}, X.-W., {Barlow}, M.~J., {Zhang}, Y., {Bastin}, R.~J., \& {Storey}, P.~J.
  2006, MNRAS, 368, 1959

\bibitem[{{Liu} {et~al.}(2001){Liu}, {Luo}, {Barlow}, {Danziger}, \&
  {Storey}}]{liuetal01}
{Liu}, X.-W., {Luo}, S.-G., {Barlow}, M.~J., {Danziger}, I.~J., \& {Storey},
  P.~J. 2001, MNRAS, 327, 141

\bibitem[{{Liu} {et~al.}(1995){Liu}, {Storey}, {Barlow}, \&
  {Clegg}}]{liuetal95}
{Liu}, X.-W., {Storey}, P.~J., {Barlow}, M.~J., \& {Clegg}, R.~E.~S. 1995,
  MNRAS, 272, 369

\bibitem[{{Liu} {et~al.}(2000){Liu}, {Storey}, {Barlow}, {Danziger}, {Cohen},
  \& {Bryce}}]{liuetal00}
{Liu}, X.-W., {Storey}, P.~J., {Barlow}, M.~J., {Danziger}, I.~J., {Cohen}, M.,
  \& {Bryce}, M. 2000, MNRAS, 312, 585

\bibitem[{{Liu} {et~al.}(2004){Liu}, {Liu}, {Barlow}, \& {Luo}}]{yliuetal04b}
{Liu}, Y., {Liu}, X.-W., {Barlow}, M.~J., \& {Luo}, S.-G. 2004, MNRAS, 353,
  1251

\bibitem[{{L\'opez-S\'anchez} {et~al.}(2007){L\'opez-S\'anchez}, {Esteban},
  {Garc{\'{\i}}a-Rojas}, {Peimbert}, \&
  {Rodr{\'{\i}}guez}}]{lopezsanchezetal06}
{L\'opez-S\'anchez}, A.~R., {Esteban}, C., {Garc{\'{\i}}a-Rojas}, J.,
  {Peimbert}, M., \& {Rodr{\'{\i}}guez}, M. 2007, ApJ, 656, 168

\bibitem[{{Ma{\'{\i}}z-Apell{\'a}niz}
  {et~al.}(2004){Ma{\'{\i}}z-Apell{\'a}niz}, {Walborn}, {Galu{\'e}}, \&
  {Wei}}]{maizapellanizetal04}
{Ma{\'{\i}}z-Apell{\'a}niz}, J., {Walborn}, N.~R., {Galu{\'e}}, H.~{\'A}., \&
  {Wei}, L.~H. 2004, ApJS, 151, 103

\bibitem[{{Mathis} {et~al.}(1998){Mathis}, {Torres-Peimbert}, \&
  {Peimbert}}]{mathisetal98}
{Mathis}, J.~S., {Torres-Peimbert}, S., \& {Peimbert}, M. 1998, ApJ, 495, 328

\bibitem[{{O'Dell} {et~al.}(2003){O'Dell}, {Peimbert}, \&
  {Peimbert}}]{odelletal03}
{O'Dell}, C.~R., {Peimbert}, M., \& {Peimbert}, A. 2003, AJ, 125, 2590

\bibitem[{{Osterbrock} {et~al.}(1992){Osterbrock}, {Tran}, \&
  {Veilleux}}]{osterbrocketal92}
{Osterbrock}, D.~E., {Tran}, H.~D., \& {Veilleux}, S. 1992, ApJ, 389, 305

\bibitem[{{Peim\-bert}(1995)}]{peimbert95}
{Peim\-bert}, M. 1995, in The Analysis of Emission Lines, ed. R.~{Williams} \&
  M.~{Livio}, 165

\bibitem[{{Peimbert}(2003)}]{apeimbert03}
{Peimbert}, A. 2003, ApJ, 584, 735

\bibitem[{{Peimbert} \& {Peimbert}(2005)}]{apeimbertpeimbert05}
{Peimbert}, A. \& {Peimbert}, M. 2005, in Rev. Mexicana Astron. Astrofis. Conf.
  Ser., ed. D.~{Torres-Peimbert} \& G.~{MacAlpine}, Vol.~23, 9

\bibitem[{{Peimbert} {et~al.}(2005){Peimbert}, {Peimbert}, \&
  {Ruiz}}]{apeimbertetal05}
{Peimbert}, A., {Peimbert}, M., \& {Ruiz}, M.~T. 2005, ApJ, 634, 1056

\bibitem[{{Peimbert}(1967)}]{peimbert67}
{Peimbert}, M. 1967, ApJ, 150, 825

\bibitem[{{Peimbert} \& {Costero}(1969)}]{peimbertcostero69}
{Peimbert}, M. \& {Costero}, R. 1969, Boletin de los Observatorios Tonantzintla
  y Tacubaya, 5, 3

\bibitem[{{Peimbert} {et~al.}(2004){Peimbert}, {Peimbert}, {Ruiz}, \&
  {Esteban}}]{peimbertetal04}
{Peimbert}, M., {Peimbert}, A., {Ruiz}, M.~T., \& {Esteban}, C. 2004, ApJS,
  150, 431

\bibitem[{{Peimbert} {et~al.}(1991){Peimbert}, {Sarmiento}, \&
  {Fierro}}]{peimbertetal91}
{Peimbert}, M., {Sarmiento}, A., \& {Fierro}, J. 1991, PASP, 103, 815

\bibitem[{{Peimbert} {et~al.}(1993{\natexlab{a}}){Peimbert}, {Storey}, \&
  {Torres-Peimbert}}]{peimbertetal93}
{Peimbert}, M., {Storey}, P.~J., \& {Torres-Peimbert}, S. 1993{\natexlab{a}},
  ApJ, 414, 626

\bibitem[{{Peimbert} \& {Torres-Peimbert}(1977)}]{peimberttorrespeimbert77}
{Peimbert}, M. \& {Torres-Peimbert}, S. 1977, MNRAS, 179, 217

\bibitem[{{Peimbert} {et~al.}(1993{\natexlab{b}}){Peimbert}, {Torres-Peimbert},
  \& {Dufour}}]{peimbertetal93b}
{Peimbert}, M., {Torres-Peimbert}, S., \& {Dufour}, R.~J. 1993{\natexlab{b}},
  ApJ, 418, 760

\bibitem[{{Peimbert} {et~al.}(1995){Peimbert}, {Torres-Peimbert}, \&
  {Luridiana}}]{peimbertetal95}
{Peimbert}, M., {Torres-Peimbert}, S., \& {Luridiana}, V. 1995, Rev. Mexicana
  Astron. Astrofis., 31, 131

\bibitem[{{Peimbert} {et~al.}(1992){Peimbert}, {Torres-Peimbert}, \&
  {Ruiz}}]{peimbertetal92}
{Peimbert}, M., {Torres-Peimbert}, S., \& {Ruiz}, M.~T. 1992, Rev. Mexicana
  Astron. Astrofis., 24, 155

\bibitem[{{Robertson-Tessi} \& {Garnett}(2005)}]{robertsontessigarnett05}
{Robertson-Tessi}, M. \& {Garnett}, D.~R. 2005, ApJS, 157, 371

\bibitem[{{Rodr{\'{\i}}guez} \& {Rubin}(2005)}]{rodriguezrubin05}
{Rodr{\'{\i}}guez}, M. \& {Rubin}, R.~H. 2005, ApJ, 626, 900

\bibitem[{{Rola} \& {Stasi{\'n}ska}(1994)}]{rolastasinska94}
{Rola}, C. \& {Stasi{\'n}ska}, G. 1994, A\&A, 282, 199

\bibitem[{{Rubin} {et~al.}(2002){Rubin}, {Bhatt}, {Dufour}, {Buckalew},
  {Barlow}, {Liu}, {Storey}, {Balick}, {Ferland}, {Harrington}, \&
  {Martin}}]{rubinetal02}
{Rubin}, R.~H., {Bhatt}, N.~J., {Dufour}, R.~J., {Buckalew}, B.~A., {Barlow},
  M.~J., {Liu}, X.-W., {Storey}, P.~J., {Balick}, B., {Ferland}, G.~J.,
  {Harrington}, J.~P., \& {Martin}, P.~G. 2002, MNRAS, 334, 777

\bibitem[{{Rubin} {et~al.}(2003){Rubin}, {Martin}, {Dufour}, {Ferland},
  {Blagrave}, {Liu}, {Nguyen}, \& {Baldwin}}]{rubinetal03}
{Rubin}, R.~H., {Martin}, P.~G., {Dufour}, R.~J., {Ferland}, G.~J., {Blagrave},
  K.~P.~M., {Liu}, X.-W., {Nguyen}, J.~F., \& {Baldwin}, J.~A. 2003, MNRAS,
  340, 362

\bibitem[{{Ruiz} {et~al.}(2003){Ruiz}, {Peimbert}, {Peimbert}, \&
  {Esteban}}]{ruizetal03}
{Ruiz}, M.~T., {Peimbert}, A., {Peimbert}, M., \& {Esteban}, C. 2003, ApJ, 595,
  247

\bibitem[{{Simpson} {et~al.}(1995){Simpson}, {Colgan}, {Rubin}, {Erickson}, \&
  {Haas}}]{simpsonetal95}
{Simpson}, J.~P., {Colgan}, S.~W.~J., {Rubin}, R.~H., {Erickson}, E.~F., \&
  {Haas}, M.~R. 1995, ApJ, 444, 721

\bibitem[{{Stasi{\' n}ska} {et~al.}(2007){Stasi{\' n}ska}, {Tenorio-Tagle},
  {Rodr\'{\i}guez}, \& {Henney}}]{stasinskaetal07}
{Stasi{\' n}ska}, G., {Tenorio-Tagle}, G., {Rodr\'{\i}guez}, M., \& {Henney},
  W.~J. 2007, A\&A, arXiv:astro

\bibitem[{{Stasi{\'n}ska} \& {Szczerba}(2001)}]{stasinskaszczerba01}
{Stasi{\'n}ska}, G. \& {Szczerba}, R. 2001, A\&A, 379, 1024

\bibitem[{{Tenorio-Tagle}(1996)}]{tenoriotagle96}
{Tenorio-Tagle}, G. 1996, AJ, 111, 1641

\bibitem[{{Torres-Peimbert} \& {Peimbert}(1977)}]{torrespeimbertpeimbert77}
{Torres-Peimbert}, S. \& {Peimbert}, M. 1977, Rev. Mexicana Astron. Astrofis.,
  2, 181

\bibitem[{{Torres-Peimbert} {et~al.}(1980){Torres-Peimbert}, {Peimbert}, \&
  {Daltabuit}}]{torrespeimbertetal80}
{Torres-Peimbert}, S., {Peimbert}, M., \& {Daltabuit}, E. 1980, ApJ, 238, 133

\bibitem[{{Tsamis} {et~al.}(2003){Tsamis}, {Barlow}, {Liu}, {Danziger}, \&
  {Storey}}]{tsamisetal03}
{Tsamis}, Y.~G., {Barlow}, M.~J., {Liu}, X.-W., {Danziger}, I.~J., \& {Storey},
  P.~J. 2003, MNRAS, 338, 687

\bibitem[{{Tsamis} {et~al.}(2004){Tsamis}, {Barlow}, {Liu}, {Storey}, \&
  {Danziger}}]{tsamisetal04}
{Tsamis}, Y.~G., {Barlow}, M.~J., {Liu}, X.-W., {Storey}, P.~J., \& {Danziger},
  I.~J. 2004, MNRAS, 353, 953

\bibitem[{{Tsamis} \& {P{\'e}quignot}(2005)}]{tsamispequignot05}
{Tsamis}, Y.~G. \& {P{\'e}quignot}, D. 2005, MNRAS, 364, 687

\bibitem[{{Viegas}(2002)}]{viegas02}
{Viegas}, S.~M. 2002, Rev. Mexicana Astron. Astrofis. Conf. Ser., 12, 219

\bibitem[{{Walter} {et~al.}(1992){Walter}, {Dufour}, \&
  {Hester}}]{walteretal92}
{Walter}, D.~K., {Dufour}, R.~J., \& {Hester}, J.~J. 1992, ApJ, 397, 196

\bibitem[{{Wesson} {et~al.}(2003){Wesson}, {Liu}, \& {Barlow}}]{wessonetal03}
{Wesson}, R., {Liu}, X.-W., \& {Barlow}, M.~J. 2003, MNRAS, 340, 253

\bibitem[{{Wesson} {et~al.}(2005){Wesson}, {Liu}, \& {Barlow}}]{wessonetal05}
---. 2005, MNRAS, 362, 424

\bibitem[{{Wyse}(1942)}]{wyse42}
{Wyse}, A.~B. 1942, ApJ, 95, 356

\end{thebibliography}

\end{document}